%% file: aipsamp.tex
\documentclass[journal]{IEEEtran}
\usepackage{blindtext, graphicx}
\usepackage{listings}

\usepackage{framed,multirow}

\usepackage{amssymb}
\usepackage{latexsym}

\usepackage{bm}
\usepackage{comment}
\usepackage{amsmath,amssymb,amsfonts}
\usepackage{algorithmic}
\usepackage{graphicx}
\usepackage{textcomp}
\usepackage{hhline}
\usepackage{multirow}
\usepackage{adjustbox}
\usepackage{booktabs}
\usepackage[numbers]{natbib}

\newcommand{\ie}{\textit{i}.\textit{e}.}
\newcommand{\eg}{\textit{e}.\textit{g}.}
\newcommand{\aka}{\textit{a}.\textit{k}.\textit{a}}
\newcommand{\norm}[1]{\left\lVert#1\right\rVert}
\newcommand{\minus}{\scalebox{0.75}[1.0]{\( - \)}}

\usepackage{graphicx}
\usepackage[font=small]{caption}

\usepackage[utf8]{inputenc}

\title{Conditional Generation of Medical Images via Disentangled Adversarial Inference}

\author{ \IEEEauthorblockN{Mohammad Havaei*,
         \thanks{M. Havaei is with Imagia, Montr\'eal, Canada.}}
\and
\IEEEauthorblockN{Ximeng Mao*,
         \thanks{X. Mao is with Montr\'eal Institute for Learning Algorithms (MILA), Universit\'e de Montr\'eal, Montr\'eal, Canada. This work was carried out while the author was interning at Imagia.}}
\and
\IEEEauthorblockN{Yiping Wang,
         \thanks{Y. Wang is with Imagia, Montr\'eal, Canada; and University of Victoria, Victoria, Canada.}}
\and
\IEEEauthorblockN{Qicheng Lao
         \thanks{Q. Lao is with Imagia, Montr\'eal, Canada;  Montr\'eal Institute for Learning Algorithms (MILA), Universit\'e de Montr\'eal, Montr\'eal, Canada; and West China Biomedical Big Data Center, West China Hospital of Sichuan University, Chengdu, China.}}
\thanks{This research was carried out in Imagia.}
\thanks{* Authors contributed equally.}
\thanks{Correspondence to M. Havaei (e-mail:\text{mohammad@imagia.com}) and Q. Lao (e-mail:\text{qicheng.lao@gmail.com}).}
}

\begin{document}

\maketitle
\thispagestyle{empty}
\pagestyle{empty}

\begin{abstract}
%%%
\input{abstract_long}
%%%%
\end{abstract}

%\linenumbers

%% main text
\section{Introduction}
\input{introduction}
\label{sec:introduction}

\section{Method}
\input{method}

\section{Experiments}
\input{experiments}

\section{Related work}
\input{related_work}
\section{Conclusion}
\input{conclusion} 
\input{acknowledgement}
%%Harvard
%\bibliographystyle{model2-names.bst}\biboptions{authoryear}

\bibliographystyle{plainnat}  
\bibliography{aipsamp}

\end{document}

%% file: abstract_long.tex
Synthetic medical image generation has a huge potential for improving healthcare through many applications, from data augmentation for training machine learning systems to preserving patient privacy.
Conditional Adversarial Generative Networks (cGANs) use a conditioning factor to generate images and have shown great success in recent years.
%While 
Intuitively, the information in an image can be divided into two parts: 1) \textit{content} which is presented through the conditioning vector and 2) \textit{style} which is the undiscovered information missing from the conditioning vector. 
Current practices in using cGANs for medical image generation, only use a single variable for image generation (\ie, content) and therefore, do not provide much flexibility nor control over the generated image.

In this work we propose DRAI---a dual adversarial inference framework with augmented disentanglement constraints---to learn from the image itself, disentangled representations of style and content, and use this information to impose control over the generation process. In this framework, style is learned in a fully unsupervised manner, while content is learned through both supervised learning (using the conditioning vector) and unsupervised learning (with the inference mechanism). We undergo two novel regularization steps to ensure content-style disentanglement. First, we minimize the shared information between content and style by introducing a novel application of the gradient reverse layer (GRL); second, we introduce a self-supervised regularization method to further separate information in the content and style variables. 

For evaluation, we consider two types of baselines: single latent variable models that infer a single variable, and double latent variable models that infer two variables (style and content). We conduct extensive qualitative and quantitative assessments on two publicly available medical imaging datasets (LIDC and HAM10000) and test for conditional image generation, image retrieval and style-content disentanglement. 
We show that in general, two latent variable models achieve better performance and give more control over the generated image. We also show that our proposed model (DRAI) achieves the best disentanglement score and has the best overall performance.

%% file: introduction.tex
% Introduction
\IEEEPARstart{S}{upervised} deep neural networks have shown great success in many applications, including those in medical imaging~\citep{esteva2017dermatologist, havaei2016hemis, havaei2017brain, jiang2019task}. However, previous works have demonstrated the data hungry nature of these methods. While there are a lot of medical images being scanned inside health centers everyday, there are various factors which prohibit training of large scale models capable of achieving expert level performance. Among others, these factors include patient privacy, difficulty in collecting diverse and unbiased datasets with expert level annotations~\citep{miotto2018deep,havaei2016deep}. Using synthetic data as a mean to circumvent the aforementioned challenges is a fascinating research venue for the medical imaging community. With the advent of newer algorithms and computational power, the desire to synthesize medical images that resemble real data %is not only possible but also necessary. 
is closer than ever to becoming a possibility.

Generative Adversarial Networks (GANs)~\citep{goodfellow2014generative} are generative models based on artificial neural networks which have proven successful in many applications~\citep{radford2015unsupervised,zhu2017unpaired,xu2017attngan,isola2017image,bang2018improved,zhang2017stackgan,zhang2017stackgan++, luc2016semantic}, including in the medical imaging domain~\citep{yi2019generative,huo2018synseg,yi2018sharpness,costa2017end, bayramoglu2017towards}. They can be thought of as transformation functions that map a sample from a prior distribution (\eg, the normal distribution) to a random sample from the learned data distribution ($p_{\text{model}}(\bm{x})$). %However, medical image datasets are known to have long-tail distributions with the majority of the mass around common diseases. Thus, it's very difficult to collect data from rare diseases which lie on the lower ends of the distribution tail.
However, GANs, as introduced in \citep{goodfellow2014generative}, are unconditioned generative models, and therefore, there is no control on the data being generated. 
Conditional generation can be very helpful as we can be more selective on which part of the learned distribution ($p_{\text{model}}(x)$)to generate data from.
This is particularly interesting in the medical imaging domain, where datasets are known to have long-tail distributions with the majority of the mass around common diseases. Therefore, it's very difficult to collect real patient data from rare diseases which lie on the lower ends of the distribution tail. 
In this respect, conditional generative models can be used to sample the learned data manifold in areas of interest;
 the generator can be conditioned on some factors which we care about, \eg, malignancy of a tumor, age group, ethnicity. Images generated in this way can be subsequently used for data augmentation, medical staff training, etc. 
Conditional GAN (cGAN)~\citep{mirza2014conditional} is a generative adversarial network where the model is conditioned during training by additional information in order to direct the generation process. This auxiliary information could be, in theory, any type of data, such as a class label, a set of tags, a text description, or even another image. %  can be used for conditional sampling of the learned distribution in order to generate images of interest. 
%One interesting application of cGANs is to sample the data manifold in areas of interest.  Medical image datasets are known to have long-tail distributions with the majority of the mass around common diseases. Thus, it's very difficult to collect data from rare diseases which lie on the lower ends of the distribution tail. 
%Conditional generation can be very helpful as we can be more selective on which part of the distribution to generate data from. 
%The generator can be conditioned on some factors which we care about, \eg malignancy of a tumor, age group, ethnicity. Images generated in this way can be subsequently used for data augmentation, medical staff training, etc. 

One common pitfall of cGAN is that the conditioning codes are extremely high-level and do not cover nuances of the data. For instance, the conditioning factor could be in the form of a sparse vector representing crude class information, such as class labels. Since the class information varies significantly within each class, the class label alone does not provide the freedom to control the individual factors of variation governing each class. %Take HAM10000 [ref] as an example, the dataset has $7$ classes with each class containing various class specific textures. % .... . We show in our experiments that 
 %Vanilla conditional GAN models fail to stay faithful to the conditioning labels as they do not capture the variation within each class.
 This challenge is exemplified in the medical imaging domain where insufficient label granularity is a common occurrence. We refer to the factors of variation that depend on the conditioning vector as {\it content}.

Another challenge in conditional image generation is that the image distribution also contains factors of variation that are agnostic to the conditioning code. These types of information are shared among different classes or different conditioning codes. In this work we refer to such information as {\it style}, which depending on the task, could correspond to position, orientation, location, background information, etc.  

Having distinct control over both attributes of content and style is very appealing when generating medical images. For example, for privacy concerns, %training purposes of medical residents 
we may want to refrain from using %private 
real patient images and rely on synthetically generated images instead. In such cases, having full control over the image generation is crucial for preserving patient privacy, \ie, removing patient information (style) while preserving disease information (content). Learning disentangled representation of content and style allow us to control the detailed nuances of the generation process. 

Another venue where inferring disentangled content and style can be of interest is unsupervised style or content based image retrieval, where we want the retrieved image to respect the style or the content of the query image. In this case, if there is no disentanglement of content and style, the shared information between the two variables, would result in an undesirable outcome. (see Section~\ref{ch:exp-retrieval} for experiments on this application).

In this work, we consider two types of information to preside over the image domain: $i)$ {\it content}, which refers to the information in the conditioning vector for image generation and $ii)$ {\it style}, which encompasses any information not covered by the conditioning vector. By definition, these two types of information are independent from one another and this independence criteria should be taken into account when training a model. By explicitly constraining the model to disentangle content and style, we ensure their independence and prevent information leakage between them. 
To achieve this goal, we introduce Dual Regularized Adversarial Inference (DRAI), a conditional generative model that leverages unsupervised learning and novel disentanglement constraints to learn disentangled representations of content and style, which in turn enables more control over the generation process. %To impose disentanglement, we introduce a novel application of the {\it Gradient Reverse Layer} (GRL)~\citep{GaninUAGLLML15} to minimize the shared information between content and style. We also present a new type of self-supervised regularization to further enforce disentanglement. 

%It is worth mentioning that s
Since there is no supervision on the style, we use an adversarial inference mechanism and learn to infer style in an unsupervised way from the image. For the content information however, we have access to both the conditioning vector and the image. %We can thus use supervised learning to learn content from the conditioning vector and unsupervised learning to learn content from the image.
The content can thus be learned both in a supervised way through the conditioning vector and unsupervised way from the image itself.  

%%By disentangling these two type of information in the data, we can control the generated image.
%to do so, we design an adversarial inference mechanism. The model has two components, the conditional generation and inference mechanism. 
%To facilitate disentanglement, 
We impose two novel disentanglement constraints to facilitate the separation of content and style: Firstly, we introduce a novel application of the {\it Gradient Reverse Layer} (GRL)~\citep{GaninUAGLLML15} to minimize the shared information between the two variables. Secondly, we present a new type of self-supervised regularization to further enforce disentanglement; using content-preserving transformations, we {\it attract} matching content information, while {\it repelling} different style information. 
     
An important feature of our model is that, in contrary to most conditional generation methods that require the conditioning vector at test time, our model has the flexibility to use either a conditioning vector or a content code inferred from a reference image. In addition, we also allow generating hybrid images by mixing the inferred style and content codes from multiple sources (see more details in Section~\ref{ch:hybrid}).

%In case the conditioning vector was available the model uses the conditional generation path, if instead a conditioning image was available, the model first uses the inference path to infer the content code and use that to generate the required image. 

We compare the proposed method with multiple baselines on two datasets. We show the advantage of using two latent variables to represent style and content for conditional image generation. To quantify style-content disentanglement, we introduce a disentanglement measure and show the proposed regularizations % for disentanglement 
can improve the separation of style and content information. We also demonstrate the use case of DRAI in the style or content based image retrieval. % [TODO: go through the experiment section and mention the main findings here!]
%and inturn, lead to better image retrieval as well as better looking hybrid image

%We also show inferring disentangled style and content can help both style and content based image retrieval.  
%We validate our model using two datasets; a CT dataset (LIDC) and color image [find better wording] dataset (HAM10000)
%\begin{itemize}
    %\item however, generation is controversial in medical imaging because of privacy issues. generative models have shown to overfit and memorize the training data. 
    %In other words, in the data distribution a particular class can have multiple modes and sometimes it is desirable to have control over those distinct modes.
    %mohammad: Not sure if we should be talking about modes here since we dont have evaluations for them. 
    %\item Also, there are many other factors of variation in the image distribution that are agnostic to the conditioning variables. These are the type of information which are shared among different classes or images with different conditioning codes. In this work we refer to those information as {\it style} and depending on the task they could correspond to position, orientation ... 
    
    %In chapter~\ref{ch:exp-retrieval} we show how our proposed model can help with this application. 

    %\item (Do we add this?) Another venue is  
    
%\end{itemize}

The contributions of this work can be summarized as follows:
\begin{itemize}
    %\item a new approach for anonymized image generation
    % mohammad: Im worried if we focus a lot on annonymization we need to do good experiments on them which we dont have the dataset for. 
    \item To the best of our knowledge, this is the first time disentanglement of content and style has been explored in the context of medical image generation. 
    \item We introduce a novel application of GRL that penalizes shared information between content and style in order to achieve better disentanglement. 
    \item We introduce a self-supervised regularization that encourages the model to learn independent information as content and style.
    %\item improved training stability through Mutual information regularization
    %\item balance measure between quality and disentanglement
    \item we introduce a quantitative content-style disentanglement measure that does not require any content or style labels. This is especially useful in real world scenarios where attributes contributing to content and style are not available. % and can be used to evaluate unsupervised inference of content and style.   
\end{itemize}

%[closing paragraph for the intro.]

%% file: method.tex
\label{ch:method}
Our proposed framework, DRAI, has two main components: a conditional image generation module and an inference module. In what follows, we explain each component individually. Note that the two modules are not independent since training is end-to-end. To ensure disentanglement between the two inferred variables, \ie, style and content, we impose disentanglement constraints which are also introduced in this section. 

\begin{figure*}[htbp]
	\centering
	\includegraphics[width=\textwidth]{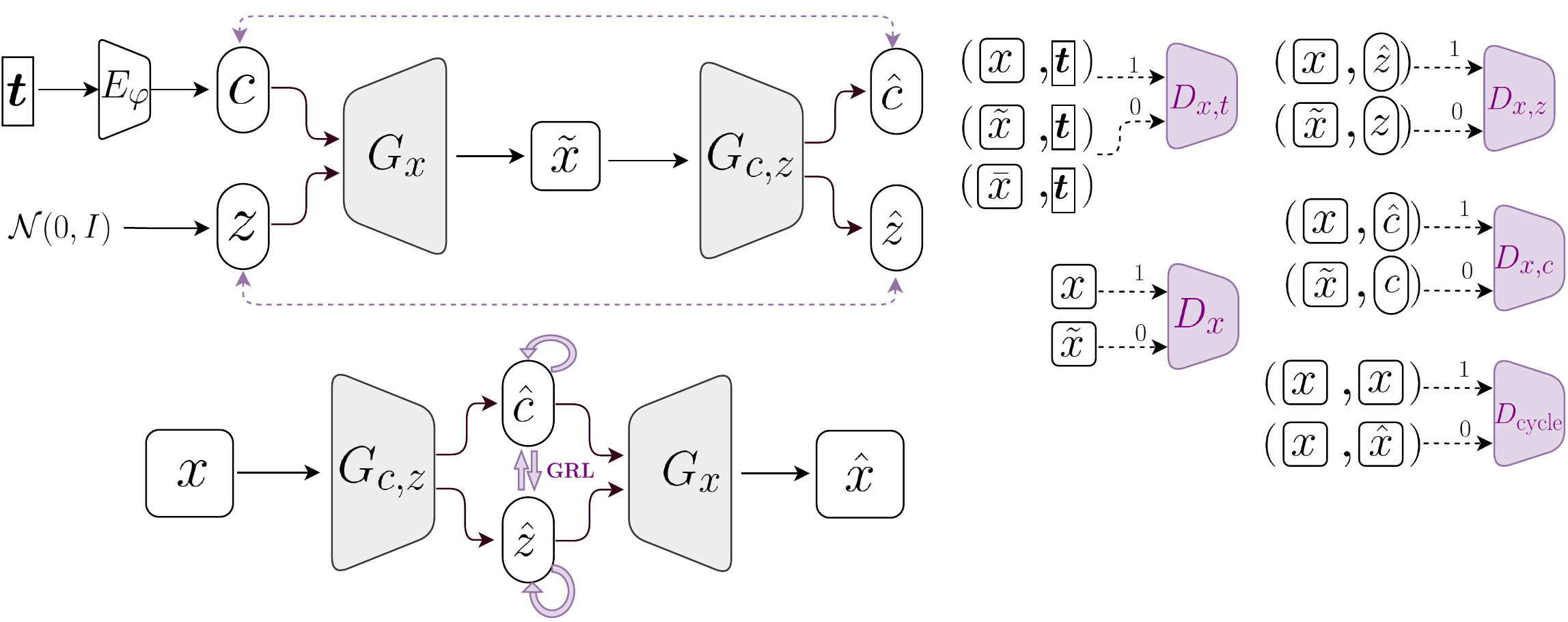}
	\caption{Overview of DRAI. The encoder $E_\varphi$ maps the conditioning vector $\bm{t}$ to a higher dimensional content space represented by the random variable $\bm{c}$ with distribution $q(\bm{c})$. The generator $G_x$ takes two inputs: a sample from the content random variable $\bm{c}$, and a sample from the style random variable $\bm{z}$ to generate $\tilde{\bm{x}}$. The dashed purple arrows mark the cycle consistency between features implemented via $\ell_1$ norm, while the solid purple arrows show the imposed disentanglement constrains. On the right hand side of the figure we show all the discriminators used for training. $\hat{\bm{c}}$ represents the inferred content, $\hat{\bm{z}}$ the inferred style, $\hat{\bm{x}}$ the reconstructed input image and $\bar{\bm{x}}$ the image with mismatched conditioning. }
	\label{fig:model}
	\vspace{-3mm}%Put here to reduce too much white space after your table 
\end{figure*}

\subsection{Overview}
We start by describing the objectives of this paper. Let $\bm{t}$ be the conditioning vector associated with image $\bm{x}$, where $\bm{t}$ could be a text description, class identity, meta data or any other piece of information for which our image generation is based on. Using the pairs $\{(\bm{t}_i,\bm{x}_i)\}, i={1, \dots , N}$, where $N$ denotes the size of the dataset, we train an inference model $G_{c,z}$ and a generative model $G_x$ such that $(i)$ the inference model $G_{c,z}$ infers content $\bm{c}$ and style $\bm{z}$ in a way that they are disentangled from each other and $(ii)$ the generator $G_x$ can generate realistic images that not only visually respect the conditioning vector $\bm{t}$ but also the style/content disentanglement. 

This framework allows for great flexibility in image generation. On one hand, conditional image generation is possible by conditioning on vector $\bm{t}$ to generate $\bm{x}$. In such cases, the content information is conveyed through the conditioning vector $\bm{t}$. On the other hand, if $\bm{t}$ is not available, we can use the inference module to infer style and content from a reference image. The inferred codes can then be used to generate images that resemble the content and/or style of the reference image. 
This is in contrast to prior works which operate on the strong assumption that the conditioning vector $\bm{t}$ is always available. 
%Thanks to the learned inference mechanism, at test time we can use the framework in a fully unsupervised manner. This means, at test time we do not need to know the content conditioning vector for $\bm{x}_a$. Instead we can infer it using the inference module explained in section ....
It is worth noting that our generative module is {\it not} constrained to require a style image. Having a probabilistic generative model allows us to sample the style code from the style prior distribution and generate images with random style attributes.% that can be learned from the data in an unsupervised manner. 

The framework also allows us to generate hybrid images by mixing style and content from various sources (details can be found in Section~\ref{ch:hybrid}). 

In the remaining of this section, we will describe each component of the proposed DRAI model. An illustration of DRAI is made in Figure~\ref{fig:model}. 
%Not only conditional generation is possible, but also  us to mix style and content from different images. More explicitly, given on a content factor $\bm{t}_a$ containing the desired content and a style image $\bm{x}_b$ containing the desired style, the framework can generate $\tilde{\bm{x}}_{ab}$ respecting the style and content from $t_a$ and $\bf{x}_b$ respectively. 

\subsection{Conditional image generation} 
%[Give a small introduction].

We follow the formulation of \citet{lao2019dual}. Our image generation module takes two vectors as input: a noise vector $\bm{z}$ (representing style) sampled from the style prior distribution and a vector $\bm{c}$ (representing content) which is an encoded representation of the conditioning vector $\bm{t}$. In order to learn richer representations for the content, we sample $\bm{c}$ from a Gaussian distribution $q_{\varphi}(\bm{c}|\bm{t}) = \mathcal{N}(\bm{\mu}_t,\,\bm{\Sigma}_t)$, where $\bm{\mu}$ and $\bm{\Sigma}$ are functions of $\bm{t}$ parameterized by a neural network with parameters $\varphi$ and are trained end-to-end with the rest of the network. To simplify the notations, we denote $q_{\varphi}(\bm{c}|\bm{t})$ as $q(\bm{c})$ throughout the paper.

In this work, we use adversarial training~\citep{goodfellow2014generative} in multiple fronts for %both generation and inference 
which we will explain in the following. 
Adversarial training was introduced by~\citep{goodfellow2014generative} where a discriminator network is trained to distinguish between real data examples (positive class) and fake examples (negative class) generated by the generator. The generator on the other hand is optimized towards fooling the discriminator. Through the adversarial game between the generator and discriminator, the distribution of the generated fake examples moves towards the distribution of real data, resulting in the generator producing realistic images. Adversarial training provides powerful implicit loss functions %(useful when it is not possible to have a closed form formulation of the loss function) 
and has shown to be very powerful in matching complex distributions. 
In order to improve the alignment between conditioning vector $\bm{t}$ and the generated image $\bm{\tilde{x}}$, we seek to match $p(\bm{\tilde{x}},\bm{t})$ with $p(\bm{x},\bm{t})$. To do so, we adopt the {matching-aware} discriminator proposed by~\citep{reed2016generative}. For this discriminator---denoted as $D_{x,t}$---the positive sample is the pair of real image and its corresponding conditioning vector $(\bm{x}, \bm{t})$, %is composed of real images and their corresponding conditioning vectors, 
whereas the negative sample pairs consist of two groups; the pair of real image with mismatched conditioning $(\bar{\bm{x}}, \bm{t})$, and the pair of synthetic image with corresponding conditioning $(G_x(\bm{z}, \bm{c}), \bm{t})$. Borrowing intuition from~\citep{belghazi2018mutual}, we can show that $D_{x,t}$ measures the mutual information between $\bm{t}$ and $\bm{x}$ and assigns higher scores to $(\bm{x},\bm{t})$ pairs with higher mutual information.  
In order to retain the fidelity of the generated images, we also train a discriminator $D_x$ that distinguishes between real and generated images. The loss function imposed by $D_{x,t}$ and $D_x$ is as follows:
\begin{multline}
    \min_G \max_D V_{\text{t2i}}(D_x, D_{x,t}, G_x) = \\
\mathbb{E}_{\bm{x} \sim p_\text{data}}[\log D_x(\bm{x})] + \mathbb{E}_{\bm{z} \sim p(\bm{z}), \bm{c} \sim q(\bm{c}})[\log(1 - D_x(G_x(\bm{z}, \bm{c})))] +\\
    \mathbb{E}_{(\bm{x}, t) \sim p_\text{data}}[\log D_{x,t}(\bm{x},t)] + \\ 
    \frac{1}{2} \big\{ \mathbb{E}_{(\bar{\bm{x}}, t) \sim p_\text{data}}[\log(1 - D_{x, t}(\bar{\bm{x}}, t))] + \\
    \mathbb{E}_{\bm{z} \sim p(\bm{z}), \bm{c} \sim q(\bm{c}), t \sim p_\text{data}}[\log(1 - D_{x, t}(G_x(\bm{z}, \bm{c}), t))] \big\}, \\[-3ex]
\end{multline}
where $\bm{\tilde{x}} = G_x(\bm{z}, \bm{c})$ is the generated image and $(\bar{\bm{x}}, t)$ designates a mis-matched pair.

\subsection{Adversarial Inference}
Latent variable models provide an efficient way to perform approximate inference in order to discover factors of variations governing the data distribution. This allows the model to reason about the data on an abstract level. While data generation is carried out through mapping the latent space $\bm{z}$ to the data space $\bm{x}$, an inference mechanism learns the inverse mapping function from $\bm{x}$ to $\bm{z}$. 
Bidirectional GAN \aka~Adversarially Learned Inference (ALI)~\citep{donahue2017adversarial,dumoulin2017adversarially} is a GAN based latent variable model that performs approximate inference by training a bidirectional discriminator to distinguish between two joint distributions: real data sample and its inferred latent code ($\bm{x}$, $\bm{\hat{z}}$) , and real latent code and its generated data sample ($\bm{z}$, $\bm{\tilde{x}}$).

%Consider two general probability distributions $q(\bm{x})$ and $p(\bm{z})$ over two domains $\bm{x} \in \mathcal{X}$ and $\bm{z} \in \mathcal{Z}$, 
\subsubsection{Single variable adversarial inference}
For a single latent variable model, let $q(\bm{x})$ represent the empirical data distribution and $p(\bm{z})$ the marginal distribution of the latent variable, specified as a simple random distribution, \eg, the standard normal distribution $\mathcal{N}(\bf{0},\bf{I})$. ALI aims to match the two joint distributions $q(\bm{x},\bm{z})=q(\bm{z}|\bm{x})q(\bm{x})$ and $p(\bm{x},\bm{z})=p(\bm{x}|\bm{z})p(\bm{z})$, which in turn implies that $q(\bm{z}|\bm{x})$ matches $p(\bm{z}|\bm{x})$. To achieve this, an encoder $G_z(\bm{x}): \hat{\bm{z}}=G_z(\bm{x}), \bm{x} \sim q(\bm{x})$ is introduced in the generation phase, in addition to the standard generator $G_x(\bm{z}): \tilde{\bm{x}}=G_x(\bm{z}), \bm{z} \sim p(\bm{z})$. The discriminator $D$ is trained to distinguish between the joint pairs $(\bm{x}, \hat{\bm{z}})$ and $(\tilde{\bm{x}}, \bm{z})$. The minimax objective of adversarial inference can be written as:
\begin{multline}\label{eq:ali_obj}
    \min_G \max_D V(D, G_x, G_z) = \\
    \mathbb{E}_{\bm{x} \sim q(\bm{x}), \hat{\bm{z}} \sim q(\bm{z}|\bm{x})}[\log D(\bm{x}, \hat{\bm{z}})] + \\
    \mathbb{E}_{\tilde{\bm{x}} \sim p(\bm{x}|\bm{z}), \bm{z} \sim p(\bm{z})}[\log(1 - D(\tilde{\bm{x}}, \bm{z}))]. \hspace{7.4mm}
\end{multline}

\subsubsection{Double variable adversarial inference}
ALI and its variants encode all the information in a single latent variable. Following \citep{lao2019dual}, we augment this framework to support two independent variables representing style ($\bm{z}$) and content ($\bm{c}$) which allows us to encode  disjoint information in each variable and ultimately disentangle style and content information. In this augmented framework, given a sample $\bm{x}$ from empirical data distribution $q(\bm{x})$, the posterior distribution over style and content is formulated as $q(\bm{z},\bm{c}|\bm{x})$. Using the adversarial inference framework,
we are interested in matching the conditional $q(\bm{z},\bm{c}|\bm{x})$ to the posterior $p(\bm{z},\bm{c}|\bm{x})$. Given the Independence assumption of $\bm{c}$ and $\bm{z}$, we have the following factorisation: 
\begin{align}
q(\bm{z},\bm{c} \mid \bm{x}) &= q(\bm{z}\mid\bm{x})q(\bm{c}\mid\bm{x}), \\
p(\bm{z},\bm{c} \mid \bm{x}) &= p(\bm{z}\mid\bm{x})p(\bm{c}\mid\bm{x}).
\end{align}

This formulation allows us to match $q(\bm{z}|\bm{x})$ to $p(\bm{z}|\bm{x})$ and $q(\bm{c}|\bm{x})$ to $p(\bm{c}|\bm{x})$, respectively. We achieve this by matching the two pairs of joint distributions:
\begin{align}
q(\bm{z},\bm{x}) &= p(\bm{z},\bm{x}), \\
q(\bm{c},\bm{x}) &= p(\bm{c},\bm{x}).
\end{align}

In the dual adversarial inference process, the feature generator $G_{c,z}$, encodes the image to infer style and content; $\hat{\bm{z}}, \hat{\bm{c}}=G_{c,z}(\bm{x}), \bm{x}\sim~q(\bm{x})$, while the image generator $G_x$ decodes samples from the style and content prior distributions into an image; $\tilde{\bm{x}}=G_x(\bm{z},\bm{c}),\bm{z}\sim~p(\bm{z}),\bm{c}~\sim~p(\bm{c})$. %To compete with $G_x$ and $G_{c,z}$, the discrimination phase also has two components:
To compete with the generators (\ie, $G_x$ and $G_{c,z}$), we train two discriminators: $D_{x,z}$ to discriminate between the pairs $(\bm{x},\hat{\bm{z}})$ and $(\tilde{\bm{x}},\bm{z})$ (sampled from $q(\bm{x},\bm{z})$ and $p(\bm{x},\bm{z})$), and $D_{x,c}$ to discriminate between the pairs $(\bm{x},\hat{\bm{c}})$ and $(\tilde{\bm{x}},\bm{c})$ (sampled from $q(\bm{x},\bm{c})$ and $p(\bm{x},\bm{c})$). %Given the above setting, the original adversarial inference objective (\ref{eq:ali_obj}) is updated as:
The dual adversarial inference objective can be thus framed as:
\begin{multline}\label{eq:dual_obj}
    \min_G \max_D V_{\text{dALI}}(D_{x,z},D_{x,c}, G_x, G_{c,z}) = \\
    % \mathbb{E}_{x \sim q(x), \hat{z}, \hat{c} \sim q(z,c|x)}[\log(D(x, \hat{z})) + \log(D(x, \hat{c}))] + \\
    % \mathbb{E}_{\tilde{x} \sim p(x|z,c), z \sim p(z), c \sim p(c)}[\log(1 - D(\tilde{x}, z)) + \log(1 - D(\tilde{x}, c))]. \\[-3ex]
    \resizebox{0.8\linewidth}{!}{
        $\mathbb{E}_{\bm{x} \sim q(\bm{x}), \hat{\bm{z}}, \hat{\bm{c}} \sim q(\bm{z},\bm{c}|\bm{x})}[\log D_{x,z}(\bm{x}, \hat{\bm{z}}) + \log D_{x,c}(\bm{x}, \hat{\bm{c}})] +$ 
    }\\
    \resizebox{\linewidth}{!}{
        $\mathbb{E}_{\tilde{\bm{x}} \sim p(\bm{x}|\bm{z},\bm{c}), \bm{z} \sim p(\bm{z}), \bm{c} \sim p(\bm{c})}[\log(1 - D_{x,z}(\tilde{\bm{x}}, \bm{z})) + \log(1 - D_{x,c}(\tilde{\bm{x}}, \bm{c}))]$.
    }\\[-3ex]
\end{multline}
%TODO: Probably talk about cycle consistency before introducing dual infernece to avoid confusion of contribution.

\subsubsection{Image cycle-consistency}
Matching the two joint distributions alone as done in ALI, does not identify the relationship between the latent codes (\ie, $\bm{z}$ and $\bm{c}$) and the data (\ie, $\bm{x}$). This results in ALI generating realistic looking images, but having poor reconstructions. \citet{li2017alice} address this issue and describe the non-identifiability problem of ALI in the single latent variable setup. To impose correlation between the latent code $\bm{z}$ and image $\bm{x}$, they propose to incorporate a loss function which enforces cycle-consistency between data sample $\bm{x}$ and the generated image from the inferred code $\hat{\bm{z}}$. More specifically, a discriminator $\text{D}_{cycle}$ is trained to distinguish between $\bm{x}$ and its reconstruction $\bm{\hat{x}}$. 
They show that in addition to achieving better reconstructions, using cycle consistency stabilizes the training of ALI.
%TODO:make reference to cycle consistency in other places...maybe copy some material from text to image. 
%The concept of cycle-consistence has been used in many applications such as domain translation, ...
The cycle-consistency loss was introduced in~\citep{zhu2017unpaired} for unpaired image to image translation using the CycleGAN architecture and has since then been used in many different applications~\citep{almahairi2018augmented,yi2017dualgan,kim2017learning}. 

In this work, we adopt cycle-consistency in a similar way as~\citep{li2017alice}. We train a discriminator $D_{cycle}$ to distinguish between pairs $(\bm{x},\bm{x})$ and $(\bm{x},\hat{\bm{x}})$ with $\hat{\bm{x}}$ being the reconstruction for $\bm{x}$; $\hat{\bm{x}}= G_x(\hat{\bm{z}}, \hat{\bm{c}})$, where $\hat{\bm{z}}, \hat{\bm{c}} = G_{c,z}(\bm{x})$. The cycle-consistency loss is denoted by $V_{cycle}$ as follows:
\begin{multline}\label{eq:cycle_obj}
    \min_G \max_D V_{\text{image-cycle}}(D_{cycle}, G_x, G_{c,z}) = \\
    \mathbb{E}_{\bm{x} \sim q(\bm{x})}[\log D_{cycle}(\bm{x}, \bm{x})] + \\
    \mathbb{E}_{\bm{x} \sim q(\bm{x}), (\hat{\bm{z}},\hat{\bm{c}}) \sim q(\bm{z},\bm{c}|\bm{x})}[\log(1 - D_{cycle}(\bm{x}, G_x(\hat{\bm{z}}, \hat{\bm{c}})))]. \\[-3ex]
\end{multline}

We also experimented with the $\ell_1$ loss as the objective function for cycle-consistency. In practice, the adversarial cycle consistency was slightly better.

\subsubsection{Latent code cycle-consistency}
%To further preserve information from a pair of style-content codes in the generated image, we maximize the mutual information between $(\bm{c}, \bm{z})$ and $\bm{x}$ using the infoGAN~\citep{chen2016infogan} framework. 

%To do so, we infer the latent code from the generated image and apply cycle consistency between the inferred and the original codes. %we maximize the mutual information between the conditionals and the generated image. We use variational information maximization introduced in [cite infogan] to maximize $I(c,z; G(c,z))$.
To further preserve information from a pair of style-content codes in the generated image, we infer the latent code from the generated image and apply cycle consistency between the inferred and the original codes.
\begin{multline}\label{eq:featurecycle} %\tag{6}
    \min_G V_{\text{code-cycle}}(G_x, G_{c,z}) = \\
    %\mathbb{E}_{\tilde{\bm{x}} \sim p(\bm{x}|\bm{z},\bm{c}), \bm{z} \sim p(z), \bm{c} \sim p(c)}
    \mathbb{E}_{(\bm{z}',\bm{c}') \sim q(\bm{z},\bm{c}|\tilde{\bm{x}}), \bm{z} \sim p(z), \bm{c} \sim q(c)}
    [\norm{\bm{z}'-\bm{z}}+\norm{\bm{c}'-\bm{c}}], %[\log(1 - D_{cycle}(\bm{x}, G_x(\hat{\bm{z}}, \hat{\bm{c}})))]. 
    \\[-3ex]
\end{multline}
where $\tilde{\bm{x}} \sim p(\bm{x}|\bm{z},\bm{c})$.

We can show that Equation (\ref{eq:featurecycle}) maximizes the mutual information between the content variable and the generated image as well as the mutual information between the style variable and the  generated image. Let $I(\bm{c};G_x(\bm{c},\bm{z}))$ denote the mutual information between the content variable and the generated image, where
\begin{equation}~\label{eq:mi_latent}
    I(\bm{c};G_x(\bm{c},\bm{z})) = H(\bm{c}) - H(\bm{c}|\tilde{\bm{x}}).
\end{equation}
Following  \cite{agakov2004algorithm}, we define a variational lower bound on $I(\bm{c};\tilde{\bm{x}})$ by rewriting the conditional entropy in Equation (\ref{eq:mi_latent}) as:
\begin{equation} \nonumber
- H(\bm{c}|\tilde{\bm{x}}) = \mathbb{E}_{\tilde{\bm{x}} \sim p(\bm{x}|\bm{z},\bm{c})}[\log q(\bm{c}|\tilde{\bm{x}}) + D_{KL}(p(\bm{c}|\tilde{\bm{x}})||q(\bm{c}|\tilde{\bm{x}}))],
\end{equation}
and by extension:
\begin{equation} \nonumber
I(\bm{\bm{c}};\tilde{\bm{x}}) = H(\bm{c}) +\mathbb{E}_{\tilde{\bm{x}} \sim p(\bm{x}|\bm{z},\bm{c})}[\log q(\bm{c}|\tilde{\bm{x}}) + D_{KL}(q(\bm{c}|\tilde{\bm{x}})||p(\bm{c}|\tilde{\bm{x}}))]. 
\end{equation}

 Using the non-negativity of $H(\bm{c})$ and $D_{KL}(p(\bm{c}|\tilde{\bm{x}})||q(\bm{c}|\tilde{\bm{x}})))$, we obtain the following lower bound on the mutual information: 
 \begin{equation} \nonumber
I(\bm{c};\tilde{\bm{x}})\geq \mathbb{E}_{\tilde{\bm{x}} \sim p(\bm{x}|\bm{z},\bm{c})}[\log q(\bm{c}|\tilde{\bm{x}})]. 
\end{equation}
 Similar derivation can be made to show 
 \begin{equation} \nonumber
I(\bm{z};\tilde{\bm{x}})\geq \mathbb{E}_{\tilde{\bm{x}} \sim p(\bm{x}|\bm{z},\bm{c})}[\log q(\bm{z}|\tilde{\bm{x}})]. 
\end{equation}
The objective function in Equation~(\ref{eq:featurecycle}), maximizes the log likelihoods $\mathbb{E}_{\tilde{\bm{x}} \sim p(\bm{x}|\bm{z},\bm{c})}[\log q(\bm{c}|\tilde{\bm{x}})]$ and $\mathbb{E}_{\tilde{\bm{x}} \sim p(\bm{x}|\bm{z},\bm{c})}[\log q(\bm{z}|\tilde{\bm{x}})]$ and by extension the variational lower bound on the terms $I(\bm{c};\tilde{\bm{x}})$ and $I(\bm{z};\tilde{\bm{x}})$.

\subsection{Disentanglement constraints}
\citet{lao2019dual} use double variable ALI as a criterion for disentanglement. However, ALI does approximate inference and does not necessarily guarantee disentanglement between variables. To further impose disentanglement between style and content, we propose additional constrains and regularization measures. 
\subsubsection{Content-Style information minimization}
\begin{figure}[t!]
	\centering
	\includegraphics[width=0.85\columnwidth]{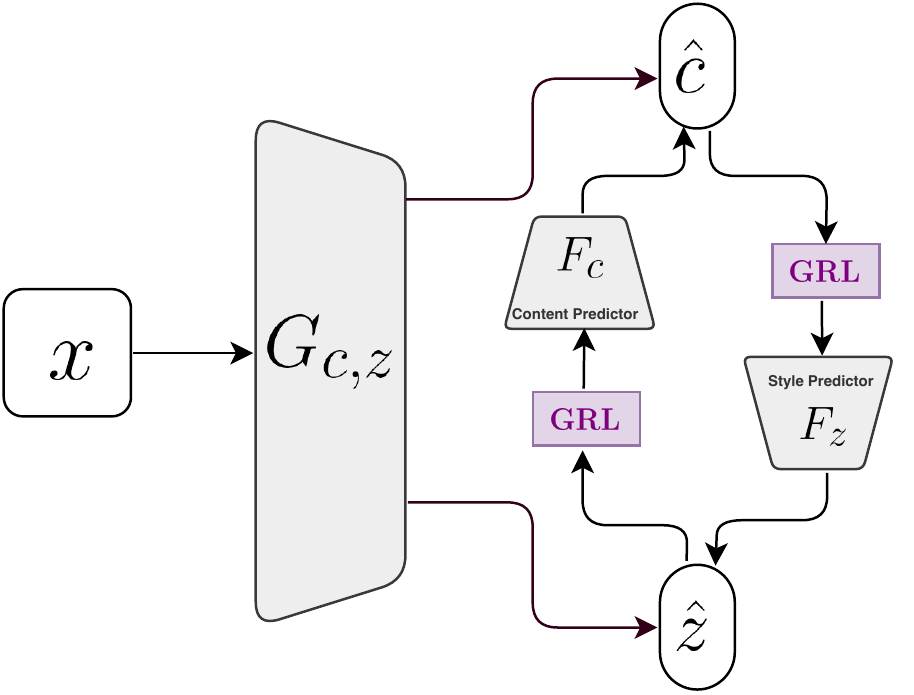}
	\caption{Content-Style information minimization. For a given image $\bm{x}$, $F_c$ is trained to predict the content $\hat{\bm{c}}$ from the style $\hat{\bm{z}}$. By reversing the direction of the gradients, the GRL penalizes $G_{c,z}$  to minimize the content information in the style variable $\bm{z}$. The same procedure is carried out to minimize style information in the content variable $\bm{c}$. }
	\label{fig:grl}
	\vspace{-3mm}%Put here to reduce too much white space after your table 
\end{figure}

The content should not include any information of the style and vice versa. We seek to \textit{explicitly} minimize the shared information between style and content. For this, we propose a novel application of
the Gradient Reversal Layer (GRL) strategy. First introduced in~\citep{GaninUAGLLML15}, the GRL strategy is used in domain adaptation methods to learn domain-agnostic features, where it acts as the identity function in the forward pass but reverses the direction of the gradients in the backward pass. In domain adaptation literature, GRL is used with a domain classifier. Reversing the direction of the gradients coming from the domain classification loss has the effect of minimizing the information between the representations and domain identity, thus, learning domain invariant features. Inspired by the literature on domain adaptation, we use GRL to minimize the information between style and content. More concretely, for a given example $\bm{x}$, we train an encoder $F_c$ to predict the content from style and use GRL to minimize the information between the two. The same process is done for predicting style from content through $F_z$, resulting in the following objective function: 
\begin{multline}\label{eq:grl}
    \min_{G}\max_F V_{\text{GRL}}(F_z, F_c, G_{c,z}) = \\
    - \mathbb{E}_{\bm{x} \sim q(\bm{x}), (\hat{\bm{z}},\hat{\bm{c}}) \sim q(\bm{z},\bm{c}|\bm{x})} [ \norm{\hat{\bm{z}}-F_z(\hat{\bm{c}})}+\norm{\hat{\bm{c}}-F_c(\hat{\bm{z}})} ]. %[\log(1 - D_{cycle}(\bm{x}, G_x(\hat{\bm{z}}, \hat{\bm{c}})))]. 
    \\[-3ex]
\end{multline}
This constrains the content feature generation to disregard style features and the style feature generation to disregard content features. Figure~\ref{fig:grl} shows a visualization of this module. 

We can show that Equation (\ref{eq:grl}) minimizes a lower bound on the mutual information between the style variable and the content variable. Here, we only provide the proof for using GRL with $F_z$ to predict style from content. Similar reasoning can be made for using GRL with $F_c$. Let $I(\bm{z};\bm{c})$ denote the mutual information between the inferred content and the style variables, where
\begin{equation}~\label{eq:mi}
    I(\bm{z};\bm{c}) = H(\bm{z}) - H(\bm{z}|\bm{c}).
\end{equation}
Once again, following  \cite{agakov2004algorithm}, we define a variational lower bound on $I(\bm{z};\bm{c})$ by rewriting the conditional entropy in (\ref{eq:mi}) as:
\begin{equation} \nonumber
\minus H(\bm{z}|\bm{c}) = \mathbb{E}_{\hat{\bm{c}}\sim q(\bm{c}|\bm{x})}[\log q(\bm{z}|\hat{\bm{c}}) + D_{KL}(p(z|\hat{\bm{c}})||q(z|\hat{\bm{c}}))]],
\end{equation}
and by extension:
\begin{equation}
I(\bm{z};\bm{c}) = H(\bm{z}) + \max_{F_z} \mathbb{E}_{\hat{\bm{c}}\sim q(\bm{c}|\bm{x})}[\log q(\bm{z}|\hat{\bm{c}})],  
\end{equation}
where the maximum is achieved when $$D_{KL}(p(\bm{z}|\hat{\bm{c}})||q(\bm{z}|\hat{\bm{c}}))] =0.$$ Since $H(\bm{z})$ is non-negative and $||\hat{\bm{z}}-F_z(\hat{\bm{c}})||$ corresponds to $\minus \log q(\bm{z}|\hat{\bm{c}})$, minimization of lower bound on mutual information can be written as:
\begin{equation}
   \min_G \max_{F_z} - \mathbb{E}_{\hat{\bm{c}}\sim q(\bm{c}|\bm{x}),\hat{\bm{z}}\sim q(\bm{z}|\bm{x})}[||\hat{\bm{z}}-F_z(\hat{\bm{c}})||],
\end{equation}
which corresponds to Equation~(\ref{eq:grl}).

\subsubsection{Self-supervised regularization}
\begin{figure}[t]
	\centering
	\includegraphics[width=0.8\columnwidth]{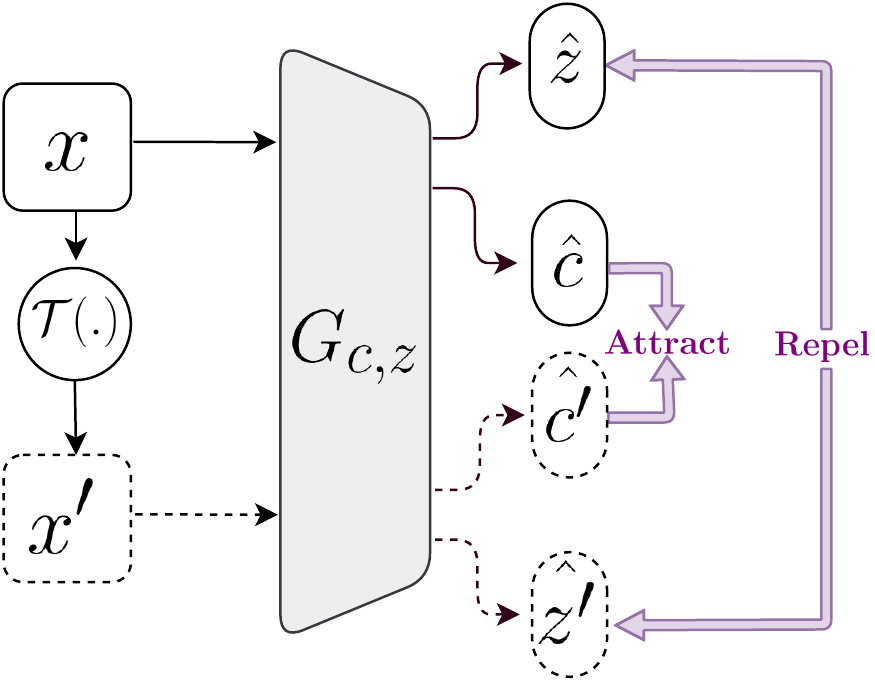}
	\caption{Self-Supervised regularization. Given $\bm{x}$ and its transformed version $\bm{x}'$, their corresponding content codes $\bm{c}$ and $\hat{\bm{c}}$ form a positive pair and the disparity between them is minimized (\ie, attract each other) while their corresponding style codes $\bm{z}$ and $\hat{\bm{z}}$ form a negative pair and the disparity between them is maximized (\ie, repel each other).}
	\label{fig:self-supervised}
	\vspace{-3mm}%Put here to reduce too much white space after your table 
\end{figure}

Self-supervised learning has shown great potential in unsupervised representation learning~\citep{oord2018representation,he2020momentum,chen2020simple}. To provide more control over the latent variables $\bm{c}$ and $\bm{z}$, we incorporate a self-supervised regularization such that the content is invariant to content-preserving transformations while the style is sensitive to such transformations. The proposed self-supervised regularization constraints the feature generator $G_{c,z}$ to encode different information for content and style. 
More formally, let $\mathcal{T}$ be a random content-preserving transformation such as a rotation, horizontal or vertical flip. For every example $\bm{x} \sim q(\bm{x})$, let $\bm{x}'$ be its transformed version; $\bm{x}'=T_i(\bm{x})$ for $T_i\sim p(\mathcal{T})$.
We would like to maximize the similarity between the inferred contents of $\bm{x}$ and $\bm{x}'$ and minimize the similarity between their inferred styles. 
This constrains the content feature generation to focus on the content of the image reflected in the conditioning vector and the style feature generation to focus on other attributes. This regularization procedure is visualized in Figure~\ref{fig:self-supervised}.
The objective function for the self-supervised regularization is defined as: 
\begin{multline}\label{eq:self}
    \min_{G} V_{\text{self}}(G_{c,z}) = 
    \mathbb{E}_{\bm{x} \sim q(\bm{x})} [\norm{\hat{\bm{c}}-\hat{\bm{c}}'} - \norm{\hat{\bm{z}}-\hat{\bm{z}}'}], %[\log(1 - D_{cycle}(\bm{x}, G_x(\hat{\bm{z}}, \hat{\bm{c}})))]. 
    \\[-3ex]
\end{multline}
where $(\hat{\bm{z}},\hat{\bm{c}}) \sim q(\bm{z},\bm{c}|\bm{x})$ and $(\hat{\bm{z}}',\hat{\bm{c}}') \sim q(\bm{z},\bm{c}|\bm{x}')$.

\subsection{Full Objective}
DRAI is a probabilistic model which requires reparameterization trick to sample from the approximate posteriors $q(\bm{z}|\bm{x})$, $q(\bm{c}|\bm{x})$ and $q(\bm{c}|\bm{t})$. We use KL divergence in order to regularize these posteriors to follow the normal distribution $\mathcal{N}(\bm{0}, \bm{I})$. 
Taking that into account, the complete objective criterion for DRAI is: 
\begin{align} \label{eq:full-obj}
\begin{split} % this is to avoid multiple equation numbering
    \min_G \max_{D, F}
    V_{\text{t2i}}(D_x, D_{x,t}, G_x)~~+\\
    V_{\text{dALI}}(D_{x,z},D_{x,c}, G_x, G_{c,z})~~+\\
    V_{\text{image-cycle}}(D_{cycle}, G_x, G_{c,z}) ~~+ \\
    V_{\text{code-cycle}}(G_x, G_{c,z}) ~~+ \\
    V_{\text{GRL}}(F_z, F_c, G_{c,z}) ~~+ \\
     V_{\text{self}}(G_{c,z}) ~~+ \\
    \lambda D_{KL}(q(\bm{z}|\bm{x}) \, || \, \mathcal{N}(\bm{0}, \bm{I})) ~~+ \\
    \lambda D_{KL}(q(\bm{c}|\bm{x}) \, || \, \mathcal{N}(\bm{0}, \bm{I})) ~~+ \\
    \lambda D_{KL}(q(\bm{c}|\bm{t}) \, || \, \mathcal{N}(\bm{0}, \bm{I})).~~~~
    \\%[-1ex]
\end{split}
\end{align} 
Figure~\ref{fig:model} illustrates various discriminators and objective functions used to train DRAI. 
\begin{figure}[t!]
	\centering
	\includegraphics[width=0.9\columnwidth]{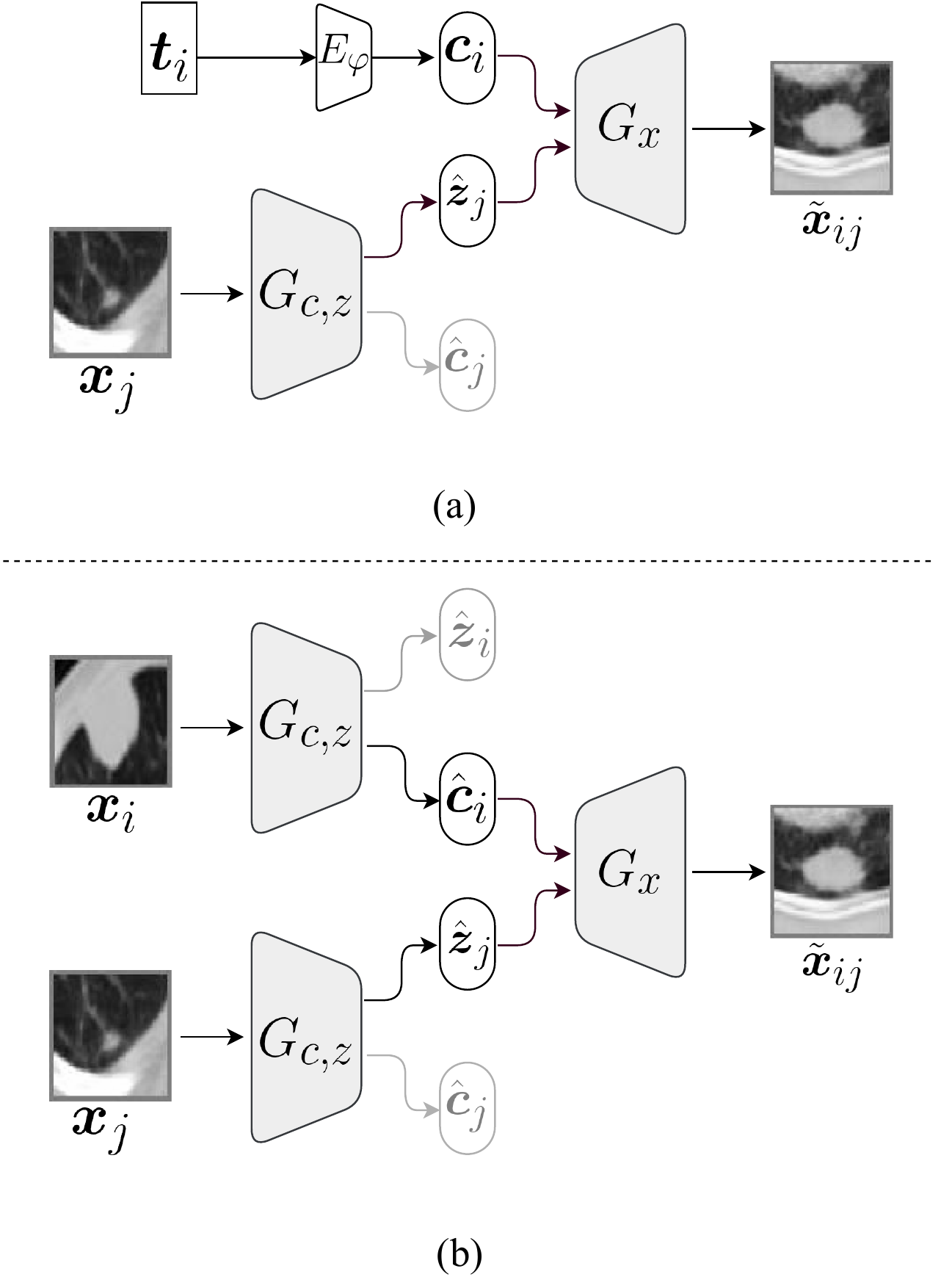}
	\caption{Hybrid image generation: (a) via the conditioning factor $\bm{t}_i$ (representing content) and the style code $\hat{\bm{z}}_j$ inferrnd from the style reference image. (b) via the content code $\hat{\bm{c}}_i$ inferred from the content reference image and the style code $\hat{\bm{z}}_j$ inferred from the style reference image.}
	\label{fig:hybrid-image}
	\vspace{-3mm}%Put here to reduce too much white space after your table 
\end{figure}

\subsection{Generating Hybrid Images} \label{ch:hybrid}
%As described earlier, having an encoder capable of inferring disentangled codes for style and content allows us to generate hybrid images where we mix style and content from different image sources. 

Thanks to our encoder that is able to infer disentangled codes for style and content and also our generator that does not have a hard constraint on requiring the conditioning embedding $\bm{t}$, we can generate hybrid images where we mix style and content from different image sources.
Let $i$ and $j$ be the indices of two different images. There are two ways in which DRAI can generate hybrid images: 

\begin{enumerate}
    \item Using a conditioning vector $\bm{t}_i$ and a style image $\bm{x}_j$:
    In this setup, we use the conditioning factor $\bm{t}_i$ as the content and the inferred $\bm{\hat{z}_j}$ from the style image $\bm{x}_j$ as the style:
\begin{align*}
\bm{c}_i = E_\varphi (\bm{t}_i) \\
\hat{\bm{z}}_j, \hat{\bm{c}}_j = G_{c,z}(\bm{x}_j) \\
\tilde{\bm{x}}_{ij} = G_x(\hat{\bm{z}}_j,\bm{c}_i).
\end{align*}

    \item Using a content image $\bm{x}_i$ and a style image $\bm{x}_j$:
    In this setup we do not rely on the conditioning factor $\bm{t}$. Instead, we infer codes for both style and content (\ie, $\hat{\bm{z}}_j$ and $\hat{\bm{c}}_i$) from style and content source images respectively. 
\begin{align*}
\hat{\bm{z}}_i, \hat{\bm{c}}_i = G_{c,z}(\bm{x}_i) \\
\hat{\bm{z}}_j, \hat{\bm{c}}_j = G_{c,z}(\bm{x}_j) \\
\tilde{\bm{x}}_{ij} = G_x(\hat{\bm{z}}_j,\hat{\bm{c}}_i)
\end{align*}
\end{enumerate}
The generation of hybrid images is graphically explained in Figure~\ref{fig:hybrid-image} for the aforementioned two scenarios.

%% file: experiments.tex
\subsection{Datasets}
We conduct experiments on two publicly available  medical imaging datasets. 
\subsubsection{HAM10000}
Human Against Machine (HAM10000) \citep{Tschandl2018TheHD}, contains approximately 10000 training images, includes 10015 dermatoscopic images of seven types of skin lesions and is widely used as a classification benchmark. One of the lesion types, ``Melanocytic nevi'' (nv), occupies around $67\%$ of the whole dataset, while the two lesion types that have the smallest data size, namely, ``Dermatofibroma'' (df) and ``Vascular skin lesions'' (vasc), have only $115$ and $143$ images respectively. Such data imbalance is undesirable for our purpose since limitations on the data size lead to severe lack of image diversity of the minority classes. For our experiments, we select the three largest skin lesion types, which in order of decreasing size are: ``nv'' with $6705$ images; ``Melanoma'' (mel) with $1113$ images; and ``Benign keratosis-like lesions'' (bkl) with $1099$ images. Patches of size $48\times48$ centered around the lesion are extracted and then resized to $64\times64$. To balance the dataset, we augment mel and bkl %for in total 
three times %and perform data augmentation 
with random flipping. We follow the train-test split provided by the dataset, and the data augmentation is done only on the training data. 
%TODO: Ximeng, can you add some text on the patch size and wheter we do any cropping of the images?

\subsubsection{LIDC} \label{ch:lidc}
The Lung Image Database Consortium image collection (LIDC-IDRI) consists of lung CT scans from 1018 clinical cases \citep{Armato2011LIDC}. In total, 7371 lesions are annotated by one to four radiologists,
 of which 2669 are given ratings on nine nodule characteristics: ``malignancy'', ``calcification'', ``lobulation'', ``margin'', ``spiculation'', ``sphericity'', ``subtlety'', ``texture'' and ``internal structure''. %We use a modified version of LIDC.  
We take the following pre-processing steps for LIDC: {\it a)} We normalize the data such that it respects the Hounsfield units (HU), {\it b)} the volume size is converted to $256\times256\times256$, {\it c)} areas around the lungs are cropped out. 
For our experiments, we extract a subset of 2D patches composing nodules with consensus from at least three radiologists. Patches of size $48\times48$ centered around the nodule are extracted and then resized to $64\times64$. Furthermore, we compute the inter-observer median of the malignancy ratings and exclude those with malignancy median of 3 (out of 5). This is to ensure a clear separation between benign and malignant classes presented in the dataset. The conditioning factor for each nodule is a 17-dimensional vector, coming from six of its characteristic ratings, as well as the nodule size. Note that ``lobulation'' and ``spiculation'' are removed due to known annotation inconsistency in their ratings~\citep{TCIALink}, and ``internal structure'' is removed since it has a very imbalanced distribution. We quantize the remaining characteristics to binary values following the same procedure of \citet{Shen2019AnID} and use the one-hot encoding to generate a 12-dimensional vector for each nodule. The remaining five dimensions are reserved for the quantization of the nodule size, ranging from $2$ to $12$ with an interval of $2$. Following the above described procedure, the nodules with case index less than $899$ are included in the training dataset while the nodules of the remaining cases are considered as the test set. By augmenting the label in such way, we exploit the richness of each nodule in LIDC, which proves to be beneficial for training.% as well, the style-content disentanglement.  

\subsection{Baselines}\label{ch:exp-baselines}
%To make these methods comparable we need to introduce dual varialbe version of them. This would make it rather odd to see them as baseline since the method we are proposing here is disentanglement through dual variable method.
To evaluate the quality of generation, inference, and disentanglement,
we consider two types of baselines. To show the effectiveness of dual variable inference, we compare our framework with single latent variable models. For this, we introduce a conditional adaptation of InfoGan~\citep{chen2016infogan} referred to as cInfoGAN and a conditional adversarial variational Autoencoder (cAVAE), both of which are explained in this section. 

To compare our approach to dual latent variable inference methods, we extend InfoGAN and cAVAE to dual variables which we denote as D-cInfoGAN and D-cAVAE respectively. 

We also compare DRAI to Dual Adversarial Inference (DAI)~\citep{lao2019dual} and show how using our proposed disentanglement constraints together with latent code cycle-consistency can significantly boost performance.
Finally, we conduct rigorous ablation studies to evaluate the impact of each component in DRAI.

%Single latent variable models where all the information is inferred in one variable

%we include two type of baseline. A single variable baselines and dual variable baselines. single variable baseline is to show the need for double variable baseline. note that one of the contributions of this paper is to use double variable models for annonymization. we can visualize that image retrieval with single variable model will mix information. quantitively we can show their inception score and FID. 
%The point of double variable models is to have a point of reference with our own model. we can show the disentanglement measure performance to justify our choice of model. 

\subsubsection{conditional InfoGAN}
InfoGAN is a variant of generative adversarial network that aims to learn unsupervised disentangled representations. In order to do so, InfoGAN modifies the original GAN in two ways. 
First, it adds an additional input $\bm{c}$ to the generator. Second, using an encoder network $Q$, it predicts $\bm{c}$ from the generated image and effectively maximizes a lower bound on the mutual information between the input code $\bm{c}$ and the generated image $\tilde{\bm{x}}$. The final objective is the combination of the original GAN objective plus that of the inferred code $\hat{\bm{c}} \sim Q(\bm{c}|\bm{x})$:
\begin{multline}
    \min_{G,Q} \max_D V_{\text{InfoGAN}}(D,G,Q) = \\
    V_{\text{GAN}}(D,G)-\lambda( \mathbb{E}_{\bm{x} \sim G(\bm{z},\bm{c}), \bm{c} \sim p(\bm{c})}[\log Q(\bm{c}|\bm{x})] + H(\bm{c})).
\end{multline}
The variable $\bm{c}$ can follow a discrete categorical distribution or a continuous distribution such as the normal distribution. InfoGAN is an unsupervised model popular for learning disentangled factors of variation~\citep{wang2019unsupervised,kurutach2018learning,ojha2019elastic}. 

\begin{figure}[t!]
	\centering
	\includegraphics[width=\columnwidth]{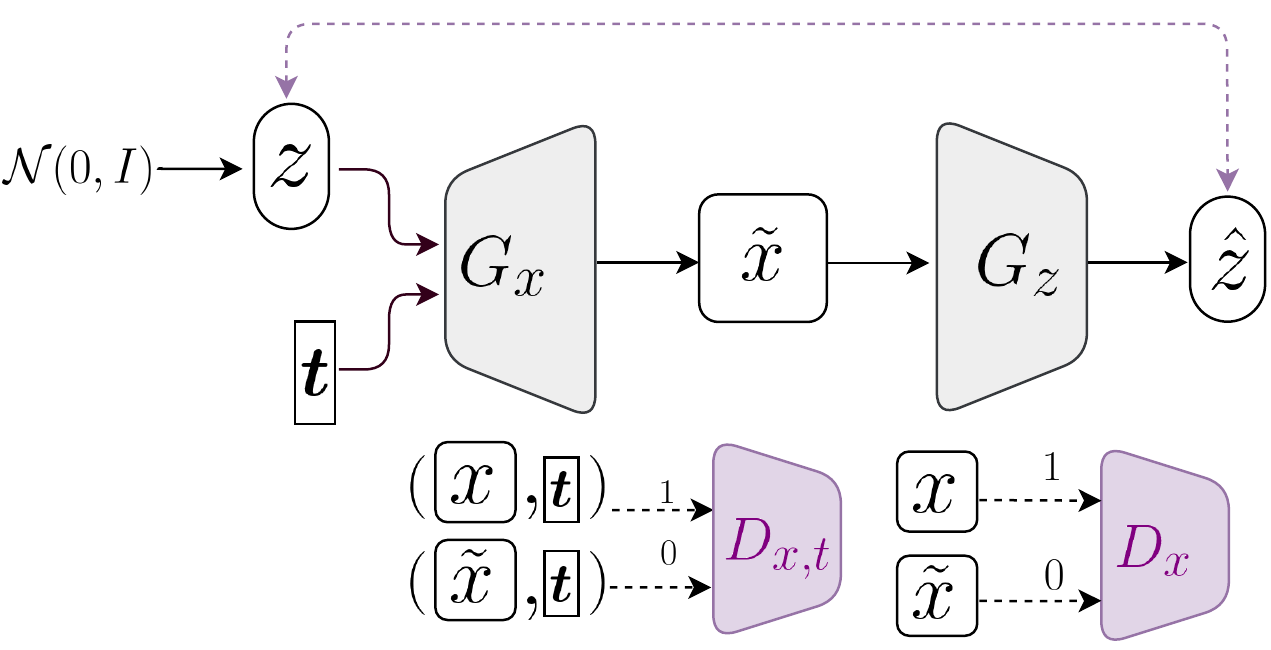}
	\caption{Conditional InfoGAN (cInfoGAN).}
	\label{fig:infogan_single}
	\vspace{-3mm}%Put here to reduce too much white space after your table 
\end{figure}

We adopt %to our needs, we present 
a conditional version of InfoGAN--denoted by cInfoGAN--which is a conditional GAN augmented with an inference mechanism using the InfoGAN formulation. We experiment with two variants of cInfoGAN; a single latent variable model (cInfoGAN) shown in Figure~\ref{fig:infogan_single}, where the discriminator $D_x$ is trained to distinguish between real ($\bm{x}$) and fake ($\tilde{\bm{x}}$) images while the discriminator $D_{x,t}$  distinguishes between the positive pair ($\bm{x}$, $\bm{t}$) and the corresponding negative pair ($\tilde{\bm{x}}$, $\bm{t}$), where $\tilde{\bm{x}} = G_x(\bm{z},\bm{t})$ and $\bm{t}$ is the conditioning vector representing content. With the help of $G_z$, InfoGAN's mutual information objective is applied on $\bm{z}$ which represents the unsupervised style. 

\begin{figure}[t!]
	\centering
	\includegraphics[width=\columnwidth]{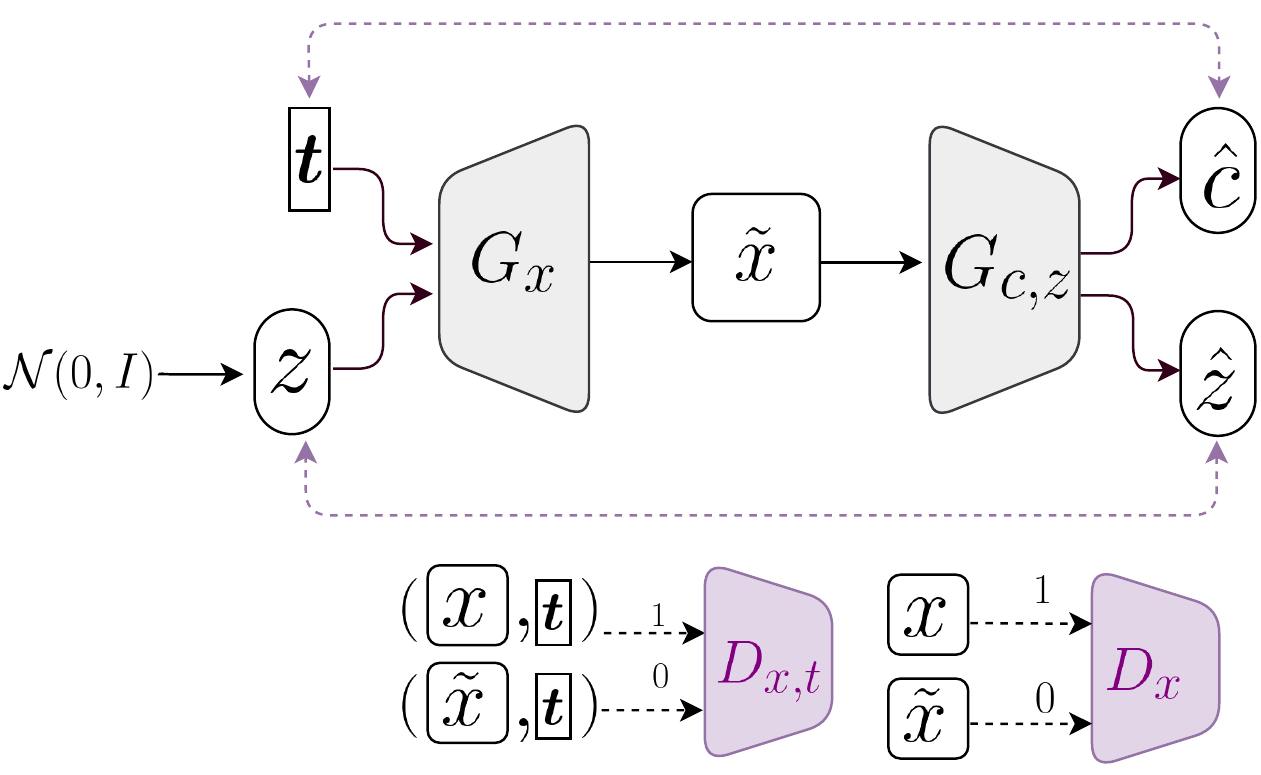}
	\caption{Dual conditional InfoGAN (D-cInfoGAN).}
	\label{fig:infogan_double}
	\vspace{-3mm}%Put here to reduce too much white space after your table 
\end{figure}

We also present a double latent variable model of InfoGAN (D-cInfoGAN) shown in Figure~\ref{fig:infogan_double} where in addition to inferring $\hat{\bm{z}}$ we also infer $\hat{\bm{c}}$ through cycle consistency using the $\ell_1$ norm.   
%[mention that in the double c infogan we maximize mutual information between z and x in a n unsupervised way by infring zhat from x generated, and use conditional gan formulation for t and x. we also empirically observed that cycle consistency between t and that helped conditional generation similar to our own model.]
\subsubsection{cAVAE}
Variational Auto-Encoders (VAEs)~\citep{kingma2014auto} are latent variable models commonly used for inferring disentangled factors of variation governing the data distribution. Let $\bm{x}$ be the random variable over the data distribution and $\bm{z}$ the random variable over the latent space. VAEs are trained by alternating between two phases, an inference phase where an encoder $G_z$ is used to map a sample from the data to the latent space and infer the posterior distribution $q(\bm{z}|\bm{x})$ and a generation phase where a decoder $G_x$ reconstructs the original image using samples of the posterior distribution with likelihood $p(\bm{x}|\bm{z})$. 

%the generation phase where a sample from the prior distribution $p(z)$ (usually an isotropic gaussian) is mapped to the data space with likelihood $p_{\theta}(x|z)$ and an inference phase, given a sample from the data distribution, the encoder inferes $q_{\phi}(z|x)$ as an approxation to the true posterior $p(z|x)$. 
VAEs maximize the evidence lower bound (ELBO) on the likelihood $p(\bm{x})$:
\begin{multline}
    \max_{G} V_{\text{VAE}}(G_x,G_z) = \\
   \mathbb{E}_{\bm{z} \sim q(\bm{z}|\bm{x})}[\log p(\bm{x}|\bm{z})]  - D_{KL}[q(\bm{z}|\bm{x}) \, || \, p(\bm{z})].
\end{multline}
%this is done by minimizing the kl divergence between $q(z)$ and the prior over the latent space $p(z)$.

\citet{kingma2014auto} also introduced a conditional version of VAE (cVAE) where $p(\bm{x}|\bm{z},\bm{c})$ is guided by both the latent code $\bm{z}$ and conditioning factor $\bm{c}$. There have also been many attempts in combining VAEs and GANs. Notable efforts are that of~\citet{larsen2016autoencoding},~\cite{mescheder2017adversarial} and~\cite{yu2019vaegan}. 

\begin{figure}[t!]
	\centering
	\includegraphics[width=0.8\columnwidth]{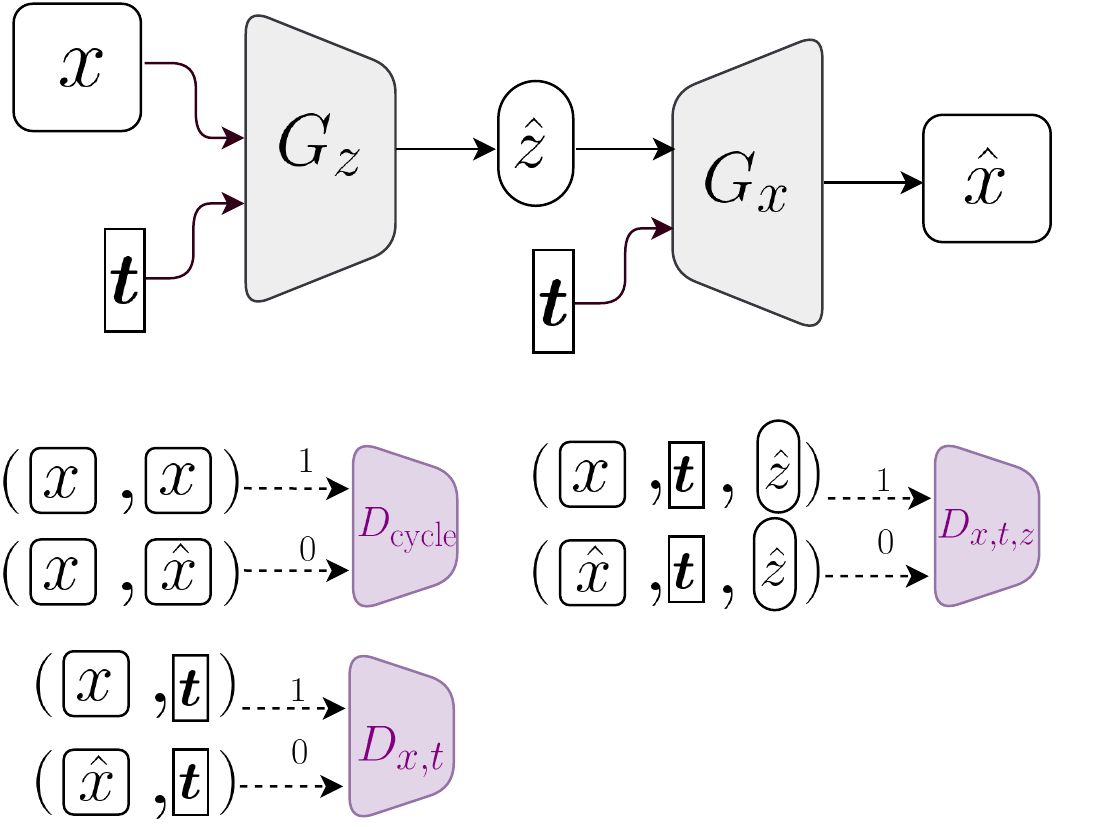}
	\caption{Conditional Adversarial VAE (cAVAE).}
	\label{fig:cbeta-vae_single}
	\vspace{-3mm}%Put here to reduce too much white space after your table 
\end{figure}

\begin{figure}[t!]
	\centering
	\includegraphics[width=0.8\columnwidth]{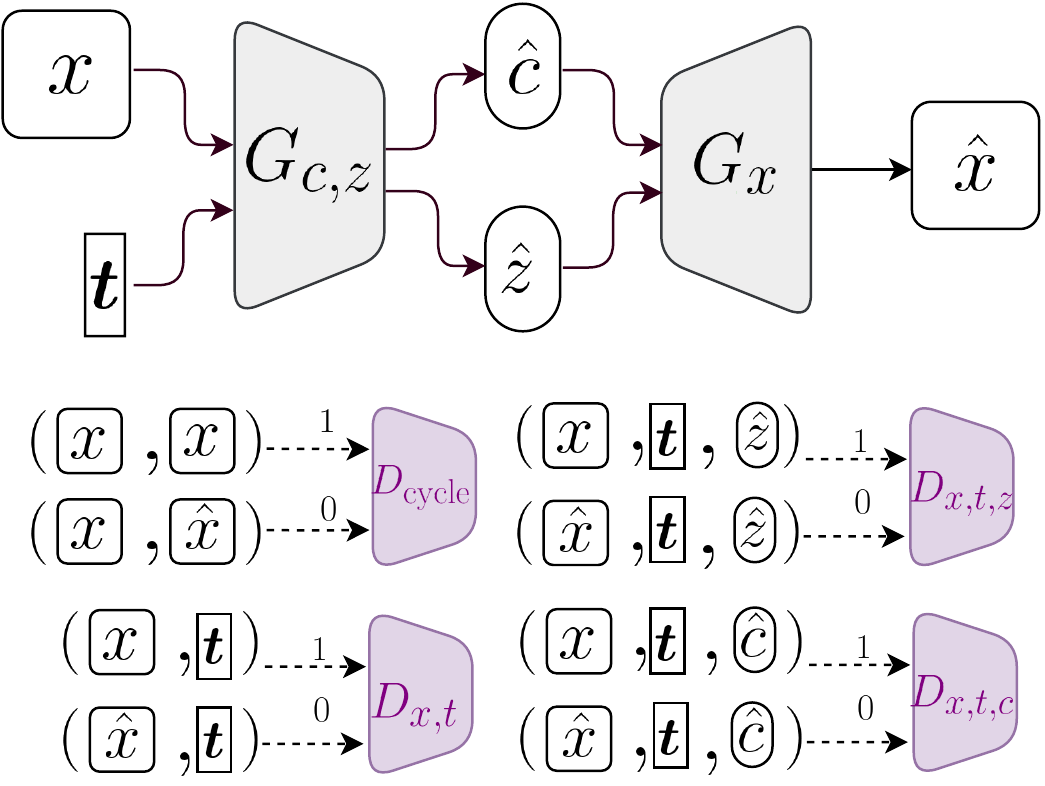}
	\caption{Dual conditional Adversarial VAE (D-cAVAE).}
	\label{fig:cbeta_vae_double}
	\vspace{-3mm}%Put here to reduce too much white space after your table 
\end{figure}

Conditional Adversarial Variational Autoencoder (cAVAE) is very similar to conditional Variational AutoEncoder (cVAE) but uses an adversarial formulation for the likelihood $p(x|z,c)$. Following the adversarial formulation for reconstruction \citep{mescheder2017adversarial,li2017alice}, a discriminator $D_{\text{cycle}}$ is trained on positive pairs ($\bm{x}$,$\bm{x}$) and negative pairs ($\bm{x}$,$\hat{\bm{x}}$), where $\hat{\bm{x}} \sim p(x|t,\hat{\bm{z}})$ and $\hat{\bm{z}} \sim q(z|x)$. For the conditional generation we train a discriminator $D_{x,t}$ on positive pairs ($\bm{x},\bm{t}$) and negative pairs ($\hat{\bm{x}},\bm{t}$), where $\bm{t}$ is the conditioning factor. We empirically discover that adding an additional discriminator $D_{x,t,z}$ which also takes advantage of the latent code $\hat{\bm{z}}$ improves inference.  
Similar to cInfoGAN, we use two versions of cAVAE: a single latent variable version denoted by cAVAE (Figure~\ref{fig:cbeta-vae_single}) and a double latent variable version D-cAVAE (Figure~\ref{fig:cbeta_vae_double}), where in addition to the style posterior  $q(\bm{z}|\bm{x})$, we also infer the content posterior $q(\bm{c}|\bm{x})$. Accordingly, to improve inference on the content variable, we add the discriminator $D_{x,t,c}$.


\subsection{Evaluation Metrics} \label{ch:exp-evaluation-metric}
We explain in detail various evaluation metrics used in our experiments. 
\subsubsection{Measure of disentanglement}
\begin{figure*}[!ht]
	\centering
	\includegraphics[width=0.8\textwidth]{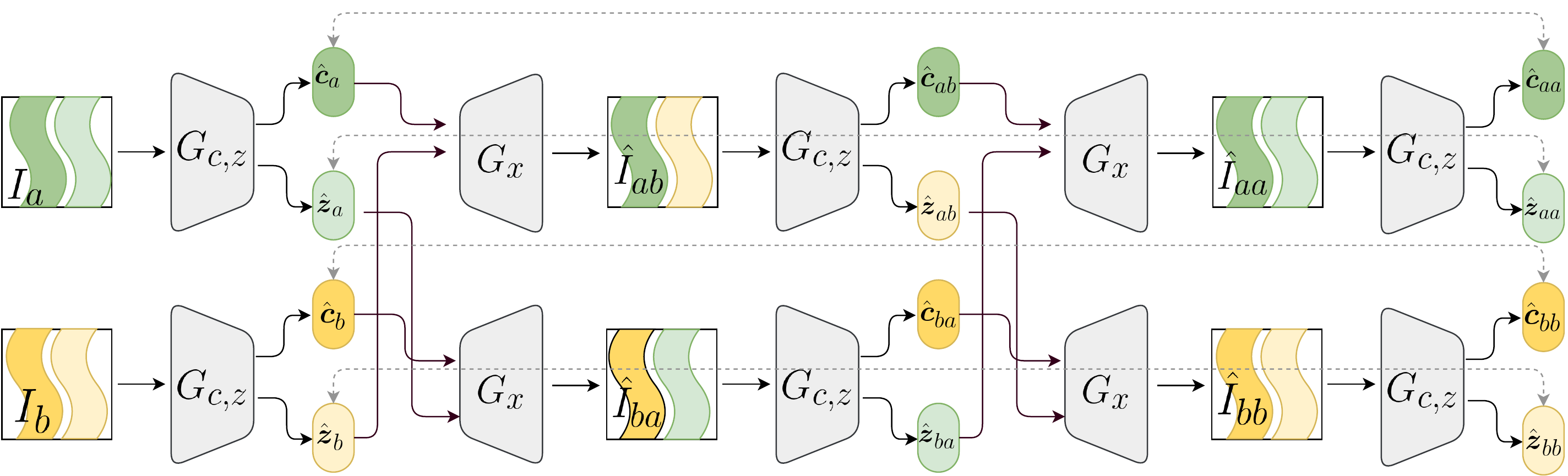}
	\caption{Cross Image Feature  Consistency (CIFC) error. CIFC is computed by first generating hybrid images of mixed style and content across two different images and then reconstructing the original images. For a more robust evaluation, CIFC is measured in the feature space.}
	\label{fig:disentanglement_measure}
	\vspace{-3mm}%Put here to reduce too much white space after your table 
\end{figure*}

Multiple methods have been proposed to measure the degree of disentanglement between variables~\citep{higgins2017beta}. %[mention some related work in this area. mention mutual information and how a tight bound on mutual information does not necessarily guarantee the desired disentanglement effect]
In this work, we propose a measure which evaluates the desired disentanglement characteristics of both the feature generator and the image generator. To have good feature disentanglement, we desire a feature generator (\ie, encoder) that separates the information in an image in two disjoint variables of style and content in such a way that $1)$ the inferred information %for both style and content 
is consistent across images. \eg, position and orientation is encoded the same way for all images; and $2)$ every piece of information is handled by {\it only} one of the two variables, meaning that the style and content variables do not share features.  
In order to measure these properties, we propose Cross Image Feature Consistency (CIFC) error where we measure the model's ability to first generate hybrid images of mixed style and content inferred from two different images and then its ability to reconstruct the original images. Figure~\ref{fig:disentanglement_measure} illustrates this process. As seen in this figure, given two images $I_a$ and $I_b$, hybrid images $I_{ab}$ and $I_{ba}$ are generated using the pairs ($\hat{\bm{c}_a}$,$\hat{\bm{z}_b}$) and ($\hat{\bm{c}_b}$,$\hat{\bm{z}_a}$) respectively. By taking another step of hybrid image generation, $I_{aa}$ and $I_{bb}$ are generated as reconstructions of $I_a$ and $I_b$ respectively. To make the evaluation robust with respect to high frequency image details, we compute the reconstruction error in the feature space. In retrospect, the disentanglement measure is computed as:
\begin{multline}
CIFC = \mathbb{E}_{(I_a,I_b) \sim q_{\text{test}}(\bm{x})} [\norm{\hat{\bm{z}}_a-\hat{\bm{z}}_{aa}}+\norm{\hat{\bm{c}}_a-\hat{\bm{c}}_{aa}}  + \\
\norm{\hat{\bm{z}}_b-\hat{\bm{z}}_{bb}}+\norm{\hat{\bm{c}}_b - \hat{\bm{c}}_{bb}}],~~~~~~~~~~~
\end{multline}
where $q_{\text{test}}(\bm{x})$ represents the empirical distribution of the test images.

\subsubsection{FID}

The Frechet inception distance (FID) score \citep{heusel2017gans} measures the distance between the real and generated data distributions. An inception model is required for calculating FID, but since the conventional inception model used for FID is pretrained on colored natural images, it is not suitable to be used with LIDC which consists of single channel CT scans. %, we can not use the existing inception model, pretrained on the RGB natural images. 
Consequently, we train an inception model on the LIDC dataset to classify benign and malignant nodules. We use InceptionV3~\citep{Szegedy2016RethinkingTI} up to layer ``\textit{mixed3}'' (initialized with pretrained ImageNet weights), and append a global average pooling layer followed by a dense layer.
\subsubsection{Inception Score}
Inception Score (IS) \citep{salimans2016improved} is another quantitative metric on image generation which is commonly used to measure the diversity of the generated images. We use the same inception model described above to calculate IS. The TensorFlow-GAN library \citep{tf-gan} is used to calculate both FID and IS.

\subsubsection{Conditional Generation Accuracy}
Since the inception model is pretrained on image labels, we can evaluate the classification accuracy of the generated images to measure the performance of the conditional generation. This metric is referred to as CGAcc through out this text. 
%The third use of the above described inception model measure the classification accuracy of the generated images. The Conditional Generation Accuracy (AGC) % insights on the amount of class determining elements present in the generated images. It 
%implicitly measures the performance of the conditional generation.
\subsubsection{Image retrieval scores}
Quantitative Image retrieval tests are conducted on LIDC dataset, where given a query image, the curated test set\footnote{See Section~\ref{ch:lidc} for details on how the data is curated} is searched to find the closest match. 

In our experiments, we construct a query dataset from the nodules in the original test set which were excluded from our curated test set (those with median malignancy equals to 3). %For both the query and reference datasets
To quantitatively evaluate the performance of image retrieval, we construct an attribution vector for each image, which consists of the conditioning vector and the nodule size. The nodule size is calculated from the nodule segmentation maps provided in the dataset. 

For each query image, the image retrieval test requires the model to find top-N nearest neighbors (we set N equal to 3 in our experiments) in the test set. The model's searching criterion is the distance between two images, which is defined by the %difference metric of 
mean absolute error (MAE) between their inferred latent representations. 

We introduce two quantitative metrics to evaluate the image retrieval performance. One is the ``disagreement divergence'' that measures the average disagreement---computed via mean square error (MSE)---between the label of the query image and those of the top-N retrieved images. % to distinguish itself from the difference metric defined above. 
 %Note that we exclude the malignancy label in the similarity score, as the malignancies of query and reference images are different by design. 
The other measure is the percentage of the label from the ground truth image found in the top-N nearest neighbors. For every query image, the ground truth image is defined as the image with the smallest disagreement divergence in the test set. %A ground truth is said to be found when at least one of the top-N nearest neighbors has sufficiently similar label to the ground truth, controlled by a predefined tolerance threshold. In other words, whenever a similarity score of less than the tolerance is recorded, the ground truth is considered as found for the current query image. 
It is important to note that since the conditioning vectors used for the evaluation of the image retrieval performance is tied with the content variable, this quantitative measure can only evaluate the content based image retrieval performance. We resort to qualitative assessments to evaluate the style based image retrieval experiments. 

\input{lidc}

\subsection{Generation evaluation}
To evaluate the quality and diversity of the generated images, we measure FID and IS (see Section~\ref{ch:exp-evaluation-metric}) for the proposed DRAI model and various double and single latent variable baselines  described in Section~\ref{ch:exp-baselines}. %The are: Dual Adversarial Inference (DAI), conditional Adversarial Variational Auto-Encoder (cAVAE), dual-conditional Adversarial Variational Auto-Encoder (dcAVAE), conditional InfoGAN (cInfoGAN) and dual conditional InfoGAN (dcInfoGAN). 
The results are reported in Table~\ref{tab:FID-IS-LIDC} and Table~\ref{tab:FID-IS-HAM} for LIDC and HAM10000 datasets respectively. 
We also report in these tables the CGAcc score which measures how well the generated images match the conditioning factors.
For the LIDC dataset, we observe all methods have comparable IS and CGAcc score while DRAI and DAI have significantly lower FID compared to other baselines, with DRAI having better performance. 
For the HAM10000 dataset, DRAI once again achieves the best FID score while D-cInfoGAN achieves the best IS. All methods seem to perform on par regarding the conditional generation accuracy CGAcc. 
It is interesting to note that compared with other baselines, DAI and DRAI achieve lower prediction intervals, which indicate more stable training process. For each model, the prediction interval is computed across four training sessions. %This is apparent from the lower prediction interval computed across multiple trained models.

\input{_ham}

\begin{figure*}[t!]
	\centering
	\includegraphics[width=0.99\textwidth]{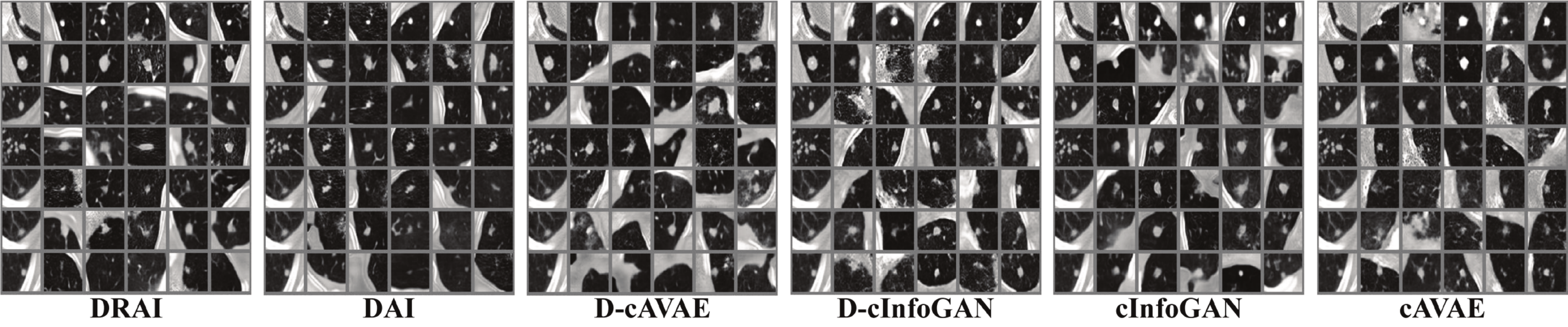}
	\caption{Conditional generations on LIDC. In every sub-graph, the first column represents the real image corresponding to the conditioning vector. The images are generated by keeping the content code ($\bm{c}$) fixed and only sampling the style codes ($\bm{z}$).}
	\label{fig:gen_lidc}
	\vspace{-3mm}%Put here to reduce too much white space after your table 
\end{figure*}
\begin{figure*}[t!]
	\centering
	\includegraphics[width=0.99\textwidth]{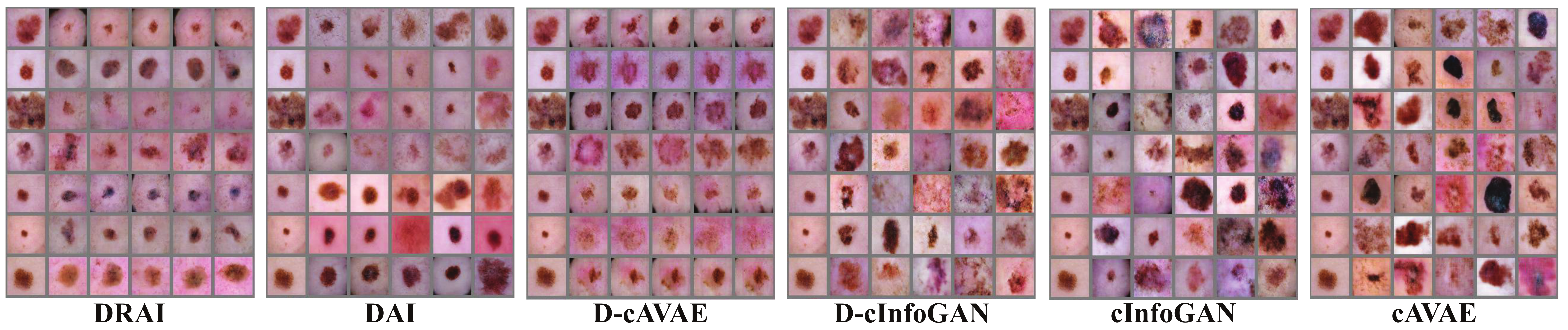}
	\caption{Conditional generations HAM10000. In every sub-graph, the first column represents the real image corresponding to the conditioning vector. The images are generated by keeping the content code ($\bm{z}$) fixed and only sampling the style codes ($\bm{z}$).}
	\label{fig:gen-ham}
	\vspace{-3mm}%Put here to reduce too much white space after your table 
\end{figure*}

We highlight that while FID and IS are the most common metrics for the evaluation of GAN based models, they do not provide the optimum assessment~\citep{barratt2018note} and thus qualitative assessment is needed. For the qualitative evaluation, we visualize samples generated by each model. We use the provided conditioning vector for the generation process and only sample the style variable $\bm{z}$. The generated samples are visualized in Figure~\ref{fig:gen_lidc} and Figure~\ref{fig:gen-ham} for LIDC and HAM10000 datasets respectively. In every sub-figure, the first column represents the reference image corresponding to the conditioning vector used for the image generation, and the remaining columns represent synthesized images. Studying these figures, we make the following observations:
\begin{itemize}
\item For LIDC, DRAI and DAI generate images with higher fidelity and less artifacts compared to other baselines. For HAM, all compared models generate realistic looking images. This could be because HAM10000 is a larger dataset. 
 \item There is evidently more content consistency in double variable models compared to single variable models. This validates the hypothesis that using two latent variables to infer style and content allows for more control over the generation process. %separation of style and content is more feasible through inferring two latent variables.
 Among double variable models, DRAI has the best content consistency. 
 \item By fixing the content and sampling the style variable, we can discover the types of information that are encoded as style and content for each dataset. We observe that the %in the case of DRAI, 
learned content information are color and lesion size for HAM10000, and nodule size for LIDC; %are kept constant 
while the learned style information are location, orientation and lesion shape for HAM10000 and background for LIDC.
We also observe that DRAI is very successful in preserving the content information when there is no stochasticity in the content variable (\ie, $\bm{c}$ is fixed).
As for other baselines, sampling style results in changing the content information of the generated images, which indicates information leak from the content variable to the style variable. The results show that compared to DAI and other baselines, DRAI achieves better separation of style and content.
\end{itemize}

\subsection{Evaluation of Style-Content Disentanglement}
Achieving good style-content disentanglement in both inference and generation phases is the main focus of this work. We conduct multiple quantitative and qualitative experiments to asses the quality of disentanglement in DRAI (our proposed method) as well as the competing baselines. 
\subsubsection{Quantitative evaluation using CIFC}
As a quantitative metric, we introduce the disentanglement error CIFC (refer to~\ref{ch:exp-evaluation-metric} for details). Table~\ref{tab:disentangle} shows results on this metric. As seen from this table, in both HAM10000 and LIDC datasets, DRAI and DAI achieve significantly lower disentanglement error compared to other baselines, indicating the importance of dual adversarial inference. 
The dual adversarial inference formulation with style and content independence assumption facilitates their separation, and allows for better disentanglement compared to D-cInfoGAN and D-cAVAE where the latent space is merely divided into two variables with separate encoders and separate discriminators. This experiment also shows that model architecture alone is not enough for style-content separation and proper objective functions are needed to guide the learning process.
%Along the same line of thought
We observe that DRAI improves over DAI by a notable margin, which demonstrates the advantage of the proposed disentanglement regularizations; on one hand, the information regularization objective through GRL minimizes the shared information between style and content variables, and on the other hand, the self-supervised regularization objective not only allows for better control of the learned features but also facilitates disentanglement. In the ablation studies (Section~\ref{ch:ablations}), we investigate the effect of the individual components of DRAI on disentanglement. 
\input{disentangle_err}

\begin{figure*}[t!]
	\centering
	\includegraphics[width=0.99\textwidth]{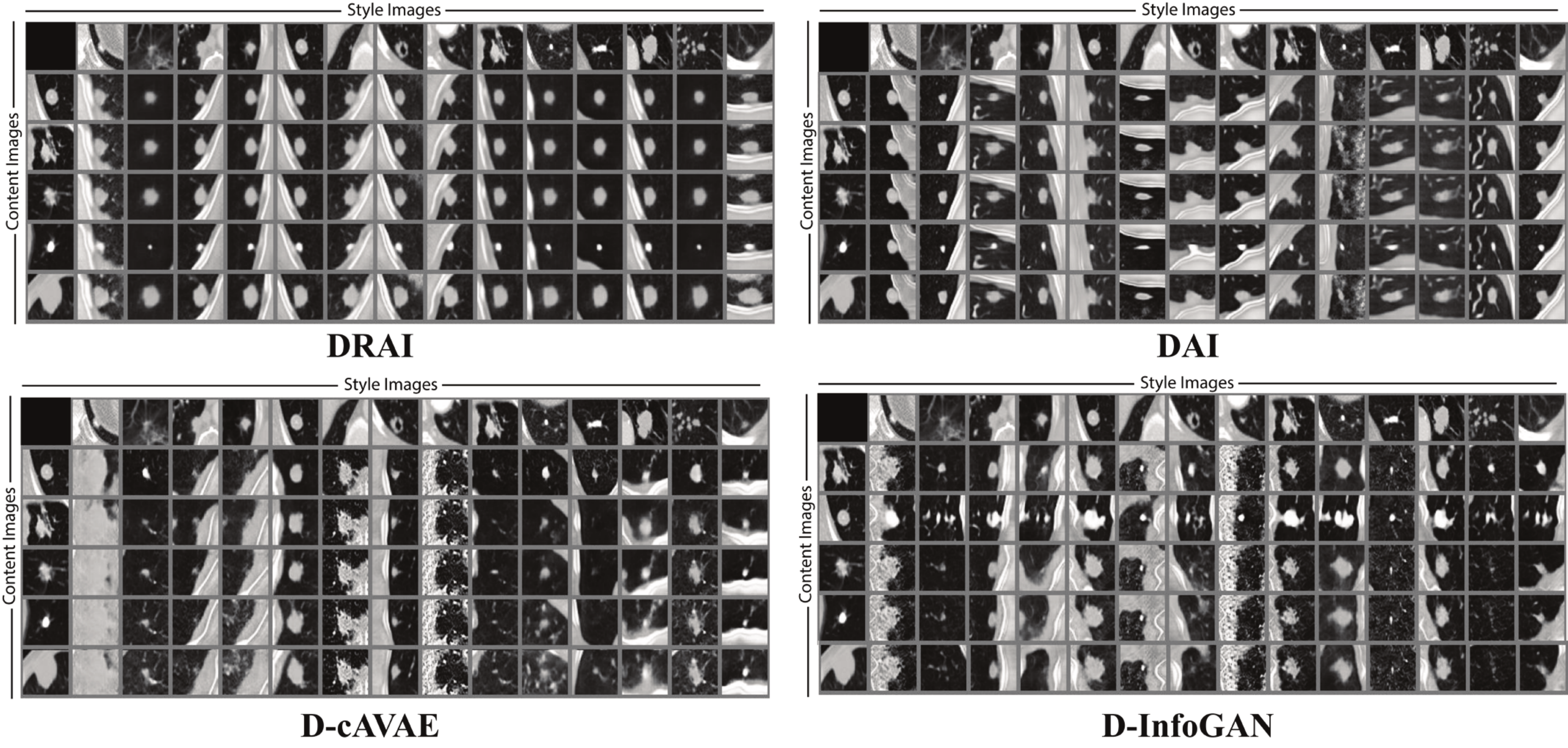}
	\caption{Qualitative evaluation of style-content disentanglement through hybrid image generation on LIDC dataset. In every sub-figure, images in the first row present style image references and those in the first column present content image references. Hybrid images are generated by using the style and content codes inferred from style and content reference images respectively.
	%infering the content code from the content reference images and style code from the style refernece images.
	}
	\label{fig:dis-lidc}
	\vspace{-3mm}%Put here to reduce too much white space after your table 
\end{figure*}

\begin{figure*}[t!]
	\centering
	\includegraphics[width=0.99\textwidth]{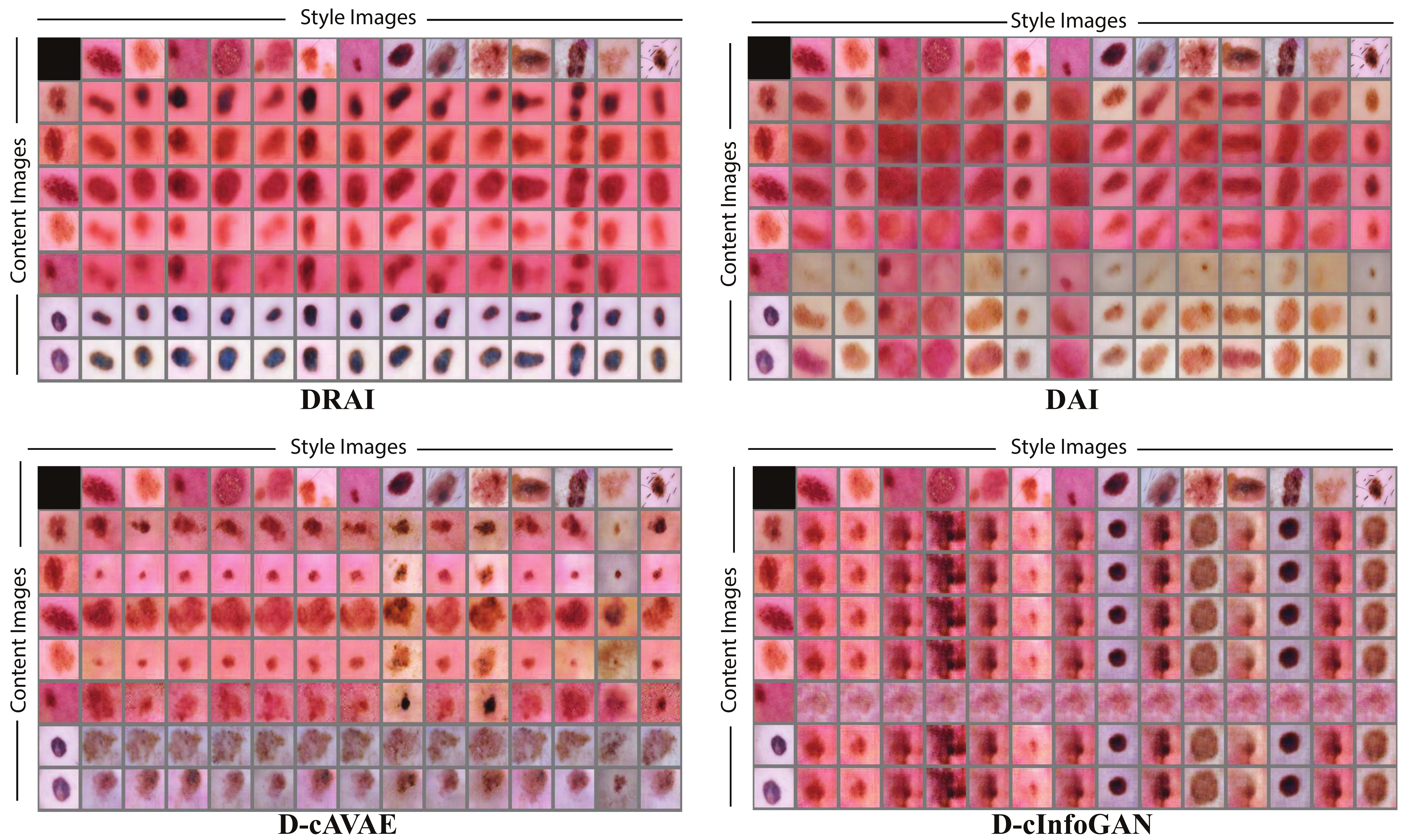}
	\caption{Qualitative evaluation of style-content disentanglement through hybrid image generation on HAM10000 dataset. In every sub-figure, images in the first row represent style image references and those in the first column represent content image references. Hybrid images are generated by using the style and content codes inferred from style and content reference images respectively.
	%infering the content code from the content reference images and style code from the style refernece images.
	}
	\label{fig:dis-ham}
	\vspace{-3mm}%Put here to reduce too much white space after your table 
\end{figure*}

\subsubsection{Qualitative evaluation}
%{\color{red} To have a more interpretable evaluation for the disentanglement, we provide three qualitative experiments: $(i)$ hybrid image generation, $(ii)$ latent space interpolation and $(iii)$ t-SNE plots for the style and content variables. }
%\paragraph{{\bf \textit{Hybrid image generation:}}}
To have a more interpretable evaluation, we qualitatively assess the style-content disentanglement through generating hybrid images by combining style and content information from different sources. We can then evaluate the extent to which the style and content of the generated images respect the corresponding style and content of the source images. Figure~\ref{fig:dis-lidc} and Figure~\ref{fig:dis-ham} show these results on the two datasets. %As shown in these figures_low the proposed model is able to better generate plausible looking hybrid images while preserving the style and content information. 
Following these figures, we make the following observations:% for DAI and DRAI models:
\begin{itemize}
\item For the LIDC dataset, DAI and DRAI learn CT image background as style and nodule as content. This is due to the fact that the nodule characteristics such as nodule size is included in the conditioning factor and thus the content tends to focus on those attributes. 
%\item In the reconstructed images, the smaller nodules often brighter (higher density?), while larger ones darker, and it could be the distributional bias presented also in the dataset.
Since the background of the CT images is not well represented in the conditioning label of the LIDC dataset, the model tends to learn features controlling the background in an unsupervised way as style. For style features, the boundary between high and low density tissues is often captured. The model also inclines towards ignoring small artifacts in the background, which suggests that it infers based on higher-level features.
\item Thanks to the added disentanglement regularizations, DRAI has the best content-style separation compared to all other baselines and demonstrates clear decoupling of the two variables. Because of the self-supervised regularization objective, DRAI assigns more emphases on capturing nodule characteristics as part of the content and background as part of the style. DAI follows closely behind but is not as successful as DRAI in inferring the correct background or nodule size. %DAI captures much less information in the content,%mostly the density, and their 
 %while the style dominates and tries to mimic the original image. % It even has nodule shape in the style, which is not desirable.
Overall, it is evident from the qualitative experiments that the proposed disentanglement regularizations help to decouple the style and content variables.
\item DRAI generates better reconstructed images compared to other baselines. The inferred content is mostly constrained to nodule size and density, which are correlated. It is likely that the nodule size and malignancy dominate the content, while other LIDC characteristics seem less discriminative in comparison. A richer and finer conditioning factor could help alleviate this.

\item For HAM10000, style encodes features shared across lesion types such as shape, orientation and location of the skin lesion, while the content constitutes lesion type, size and color information of both the lesion and background. 
%\item As HAM10000 just has class as label. The content is supposed to be a classifier, or sort. Having color in the content can be a training effect,
\item The conditioning factor for HAM10000 is only comprised of lesion type, so learning color and lesion size as part of the content, is the result of inductive biases caused by inferring {\it two} latent variables to learn style and content, and also the result of objective functions such as the {\it self supervised regularization}. The unsupervised learning of content (in addition to the supervised learning) helps bring about a richer content representation. 
%\item Style encodes common features shared across lesion types. These features include location 
\item DRAI also shows clear separation of style and content on HAM10000 dataset. DAI fails to disentangle the two vectors and some features such as color are captures by both the style and content variables. As for D-cInfoGAN and D-cAVAE models, the hybrid image generation is mostly dominated by either the style (for D-cInfoGAN) or the content (for D-cAVAE).
\item HAM10000 is a harder dataset in terms of the vast lesion variability. Using separate style and content vectors without an independence assumption between them (\eg, D-cInfoGAN and D-cAVAE), results in the coupling of the two representations which as seen in Figure~\ref{fig:dis-lidc} and Figure~\ref{fig:dis-ham}, is detrimental.% using vectors in those models would render better generation results. 
\item Requiring disentanglement adds an additional regularization effect on the generation task, which introduces a trade-off between disentanglement and generation. This is evident in the HAM10000 dataset where we observe images generated by double variable models (especially  DRAI and DAI where there is more regularization involved) are blurrier than those generated by single variable models. 
Careful tuning of the hyper-parameters is required to generate high quality images which also satisfy the disentanglement criteria for style and content.
\end{itemize}
\begin{figure*}[t!]
	\centering
	\includegraphics[width=0.99\textwidth]{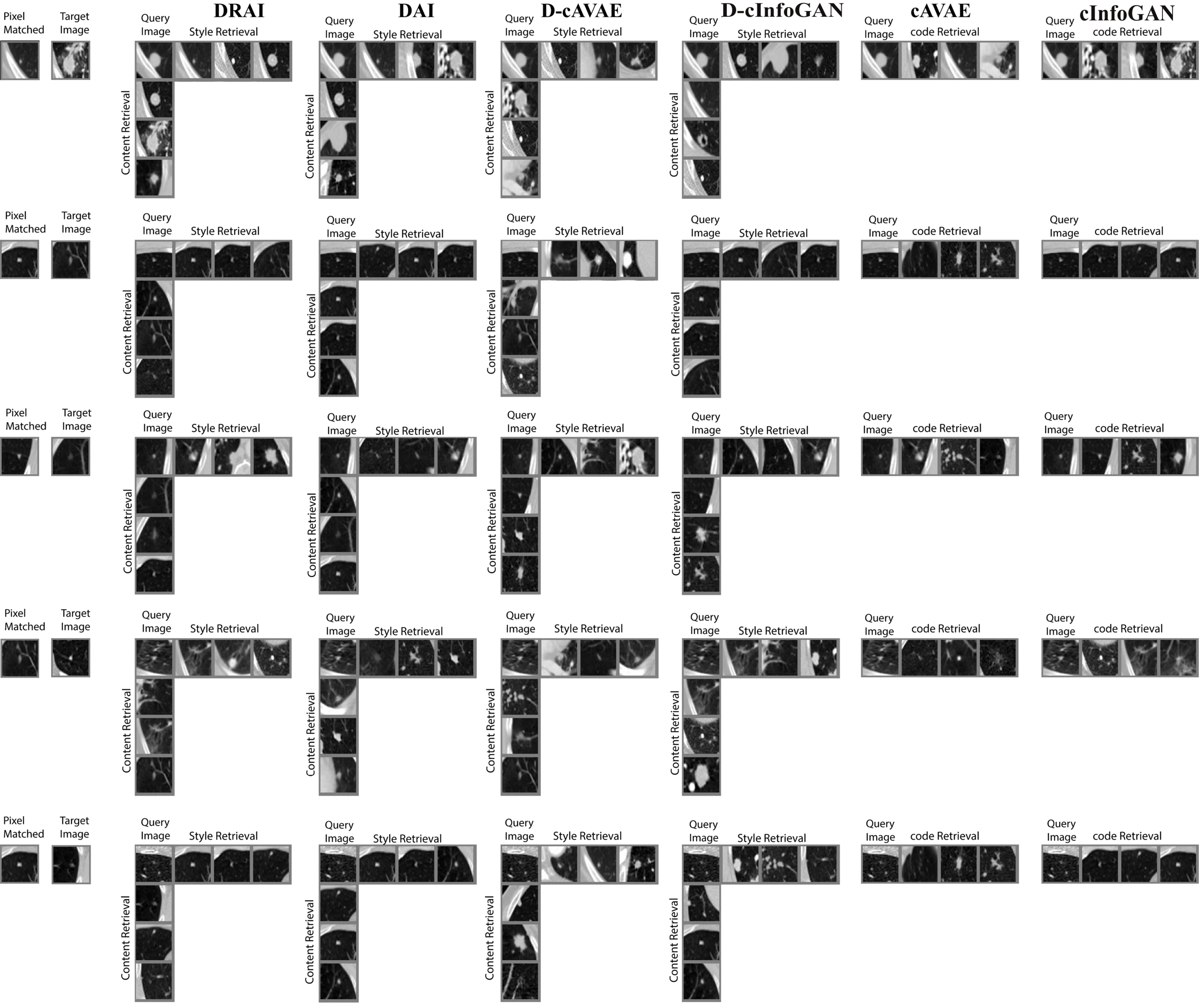}
	\caption{Style-Content image retrieval on LIDC dataset. The ``Target image'' is the image in the testset with the lowest disagreement divergence to the query image. the ``Pixel Matched'' image is the nearest neighbor to the query image in the pixel space. In every sub-figure, images to the right of the query image represent the nearest neighbors computed via the style code, while images at the bottom, represent the nearest neighbors computed via the content code.}
	\label{fig:retrieval-lidc}
	\vspace{-3mm}%Put here to reduce too much white space after your table 
\end{figure*}

\begin{figure*}[t!]
	\centering
	\includegraphics[width=0.99\textwidth]{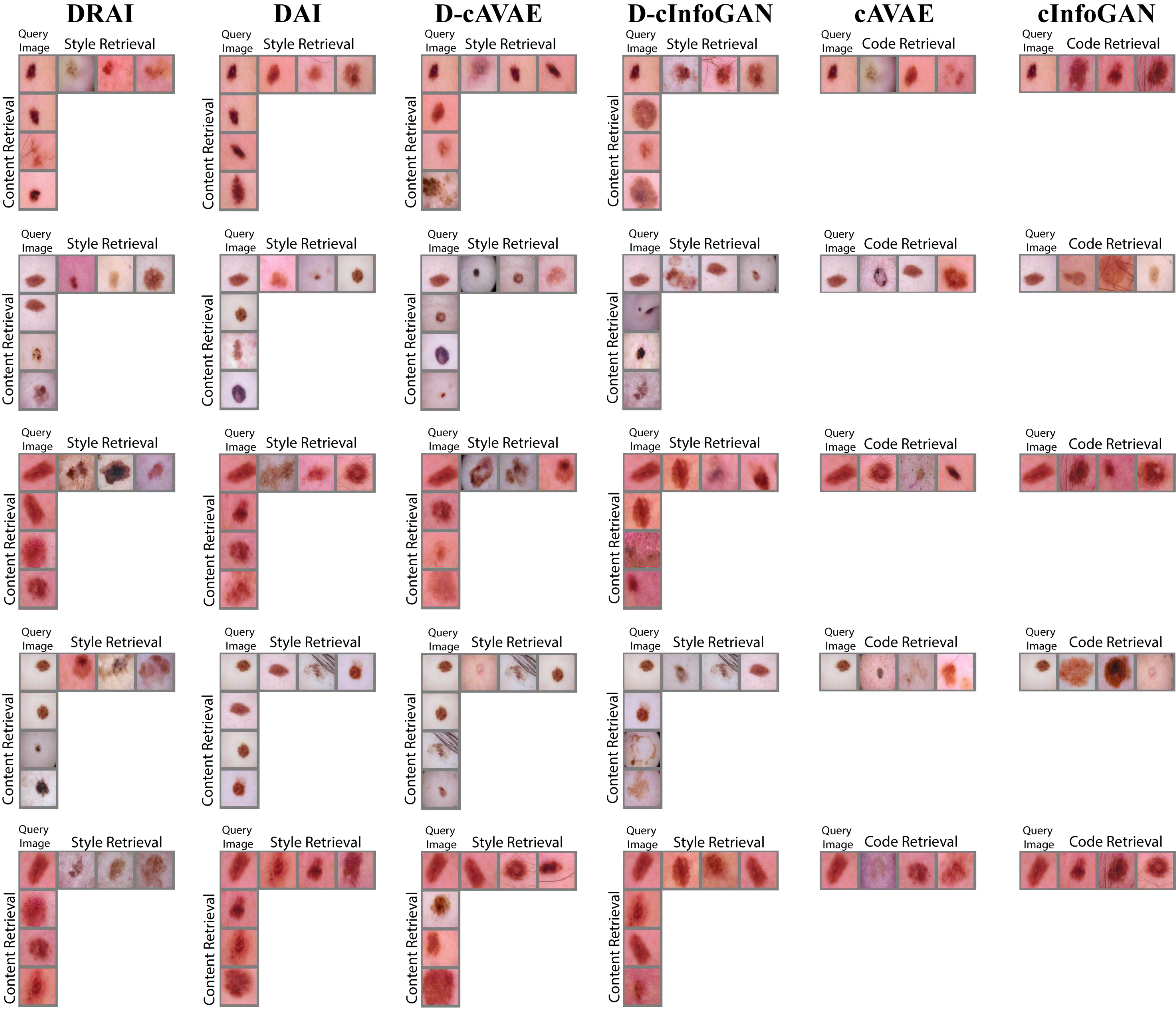}
	\caption{Style-Content image retrieval on HAM10000. In every sub-figure, images to the right of the query image represent the nearest neighbors computed via the style code, while images at the bottom, represent the nearest neighbors computed via the content code.}
	\label{fig:retrieval-ham}
	\vspace{-3mm}%Put here to reduce too much white space after your table 
\end{figure*}
\subsubsection{Image retrieval}

\label{ch:exp-retrieval}
%{\color{red} \sout{While the above experiments evaluate disentanglement on both inference and generation, we would also like to further assess} For further assessment, we could also evaluate} 
While the above experiments evaluate disentanglement on both inference and generation, we would also like to further assess how the inferred style and content features are disentangled from each other in the retrieval test. For a given query image, we can retrieve its closest matches in a reference dataset, in terms of style and content separately. The style-content disentanglement can then be evaluated by investigating the style and content of the retrieved images. %We propose style information and content information image retrieval experiments

We resort to qualitative evaluation of style based image retrieval since there is no label on style.
We could however, evaluate content based image retrieval, based on the conditioning vector (nodule size and other characteristics)
%We could however, evaluate the inferred content information based on the conditioning vector (nodule size and other characteristics) 
associated with each image. This allows us to evaluate how well each model can infer correct content information.% This could give us some hints on the style-content disentanglement since the conditioning label is only dependant on the content, and in circumstances where the content from the style image contradicts that of the content image, the contaminated inferred content will cause low score on  content based image retrieval.
 Quantitative results on content based image retrieval are presented in Table~\ref{tab:retrieval} where we present the disagreement divergence and label overlap. % for the quantitative evaluation of content based image retrieval.
As seen in this table, while DAI and DRAI have relatively similar performance, other baselines (\ie, D-cINfoGAN and D-cAVAE) significantly under perform.
Interestingly, ``Pixel Matched'', which is the nearest neighbor to the query image computed in the image space, focuses only on the background (\ie, the dominant factor in the image). This is also quantitatively shown in Table~\ref{tab:retrieval}; Pixel Matched performs poorly on both metrics.

For the quantitative evaluation, we resort to using only the LIDC dataset since it has a richer conditioning label compared to HAM10000 dataset. 

For the qualitative evaluation, the top nearest neighbors to the query image with respect to its inferred style and content codes are retrieved. 
Figure~\ref{fig:retrieval-lidc} and Figure~\ref{fig:retrieval-ham} showcase the qualitative evaluation of style based and content based image retrieval on LIDC and HAM10000 datasets respectively. In each figure, the horizontal images next to the query image represent the retrieved images based on the style information, while the vertical images represent the retrieved images based on the content information. For single variable baselines \ie,  cAVAE and cInfoGAN, image retrieval is performed based on the inferred single code. 
We observe that DRAI manages to retrieve images based on style or content without allowing interference from the other variable (\ie,  content or style respectively). As for the other baselines, since the style and content information are intertwined, the retrieved images have both style and content information which is undesirable. 
\input{retreival}

\subsection{Ablation studies} 
\label{ch:ablations}

As discussed in Section~\ref{ch:method}, our proposed method DRAI is composed of different components to ensure the quality of inference and style-content disentanglement. In this section, we perform ablation studies to evaluate the effect of each component, both quantitatively and qualitatively. Ablated models use the same architecture with the same amount of parameters. The quantitative assessment is presented in Table~\ref{tab:ablation_lidc} and Table~\ref{tab:ablation_ham} where we study the performance of each component with respect to FID, IS, CGAcc and CIFC metrics. We observe that on both LIDC and HAM10000, each added component improves over DAI, while the best performance is achieved when these components are combined together to form DRAI. 

\begin{figure*}[t!]
	\centering
	\includegraphics[width=0.99\textwidth]{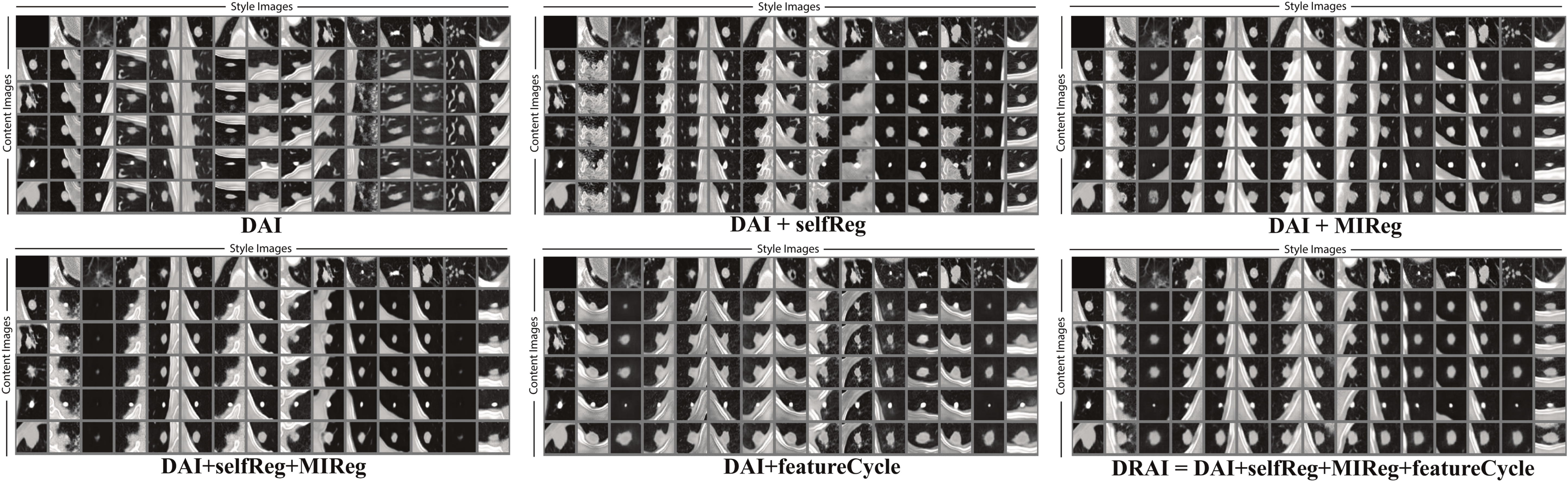}
	\caption{Qualitative ablation study of style-content disentanglement through hybrid image generation on LIDC dataset.}
	\label{fig:ablation-lidc}
	\vspace{-3mm}%Put here to reduce too much white space after your table 
\end{figure*}

\begin{figure*}[t!]
	\centering
	\includegraphics[width=0.99\textwidth]{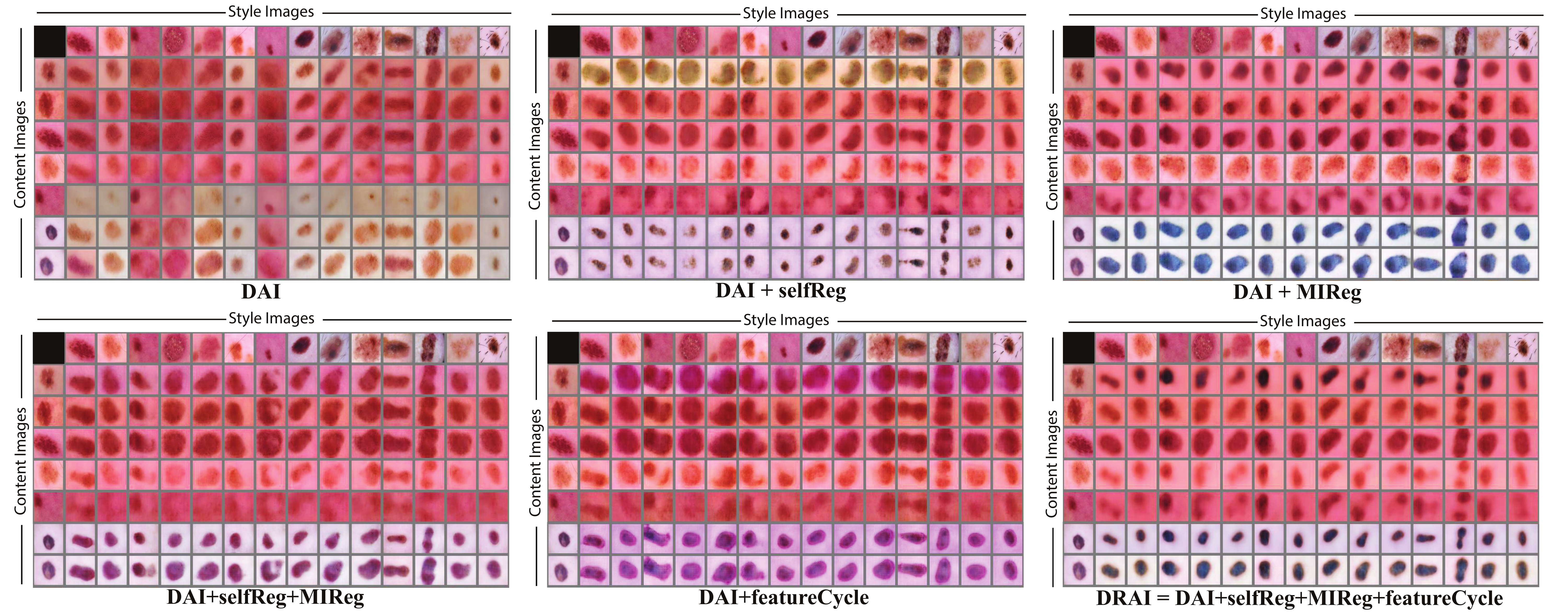}
	\caption{Qualitative ablation study of style-content disentanglement through hybrid image generation on HAM10000 dataset.}
	\label{fig:ablation-ham}
	\vspace{-3mm}%Put here to reduce too much white space after your table 
\end{figure*}

We also provide qualitative analysis of ablation studies which are visualized in Figure~\ref{fig:ablation-lidc} and Figure~\ref{fig:ablation-ham} for LIDC and HAM10000 datasets respectively. % For the LIDC dataset, we observe that DAI alone is not able to generate images which res
Our qualitative results also support the quantitative findings and show each added component in DRAI namely self-supervised regularization (``selfReg''), shared information minimization (``MIReg'') and latent code cycle consistency (``featureCycle'') can improve DAI. We observe  selfReg and MIReg have significant impact on disentanglement when combined together. However, the model may suffer from over regularization especially in the case of LIDC dataset where some content information \eg, nodule size is under estimated. This effect is alleviated when the featureCycle term is integrated. We achieve the best disentanglement with style-content preservation in DRAI where we integrate all 3 components to DAI.  

\begin{figure*}[t!]
	\centering
	\includegraphics[width=0.99\textwidth]{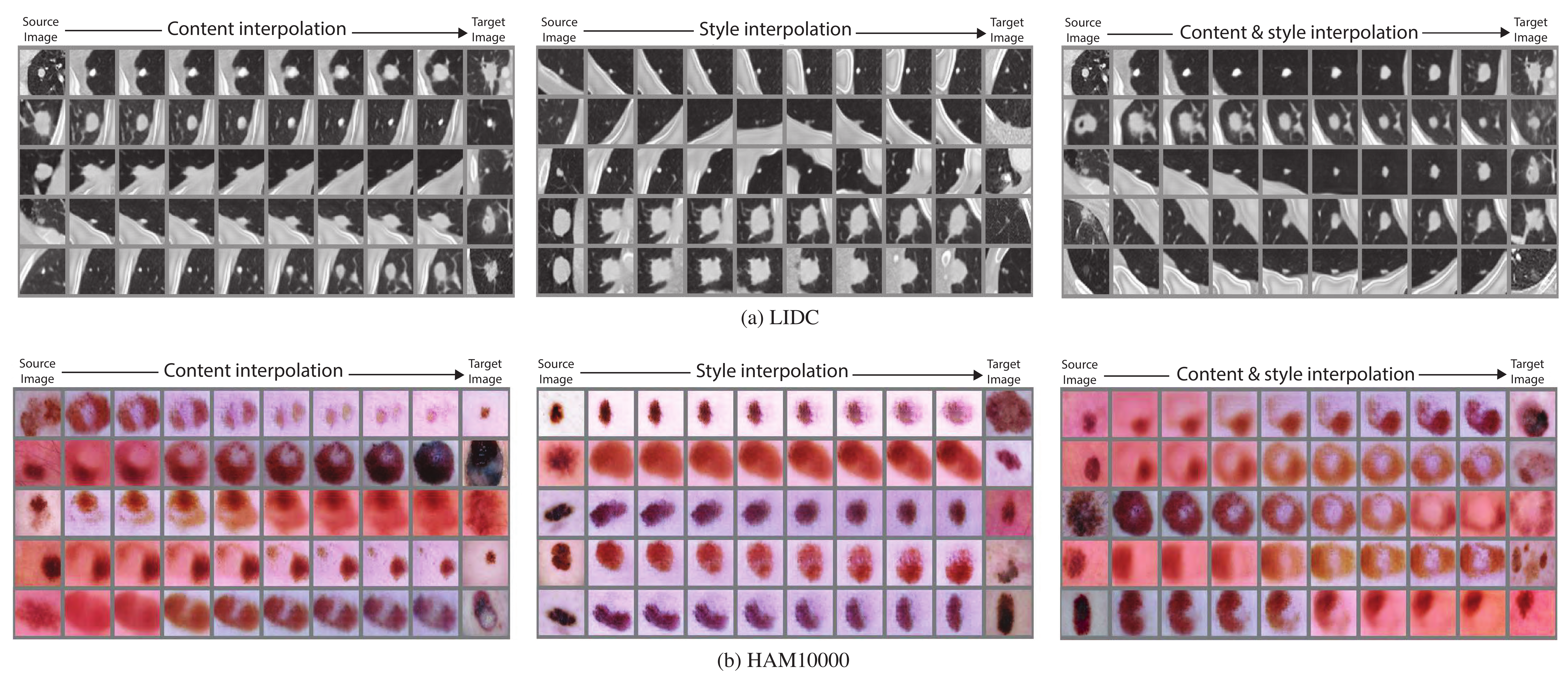}
	\caption{Latent space interpolation between testset examples for (a) LIDC and (b) HAM10000. For each dataset, we show (left) content only interpolation from source to target images while maintaining style of the source image, (middle) style only interpolation from source to target images while maintaining content of the source image and (right) both style and content interpolation. 
	}
	\label{fig:interpolation}
	\vspace{-3mm}%Put here to reduce too much white space after your table 
\end{figure*}

\paragraph{{\bf \textit{Latent space interpolation}}}
For our best performing model DRAI, as a sanity check for memorization and overfitting, we look at latent space interpolations between testset examples as shown in Figure~\ref{fig:interpolation} for both LIDC and HAM10000. We conduct 3 tests for each dataset: $(i)$ interpolating the content while maintaining the style of the source image, $(ii)$ interpolating the style while maintaining the content of the source image, $(iii)$ interpolating both style and content. 

In all 3 tests, we observe smooth transitions between pairs of examples with plausible intermediary image generations. This experiment also demonstrates the disentanglement power of DRAI; in the content interpolation experiment, DRAI maintains the style of the source image while the content gradually transitions from the source image to the target image, and in the style interpolation test, the content is maintained and the style transitions.

\begin{figure*}[t!]
	\centering
	\includegraphics[width=0.99\textwidth]{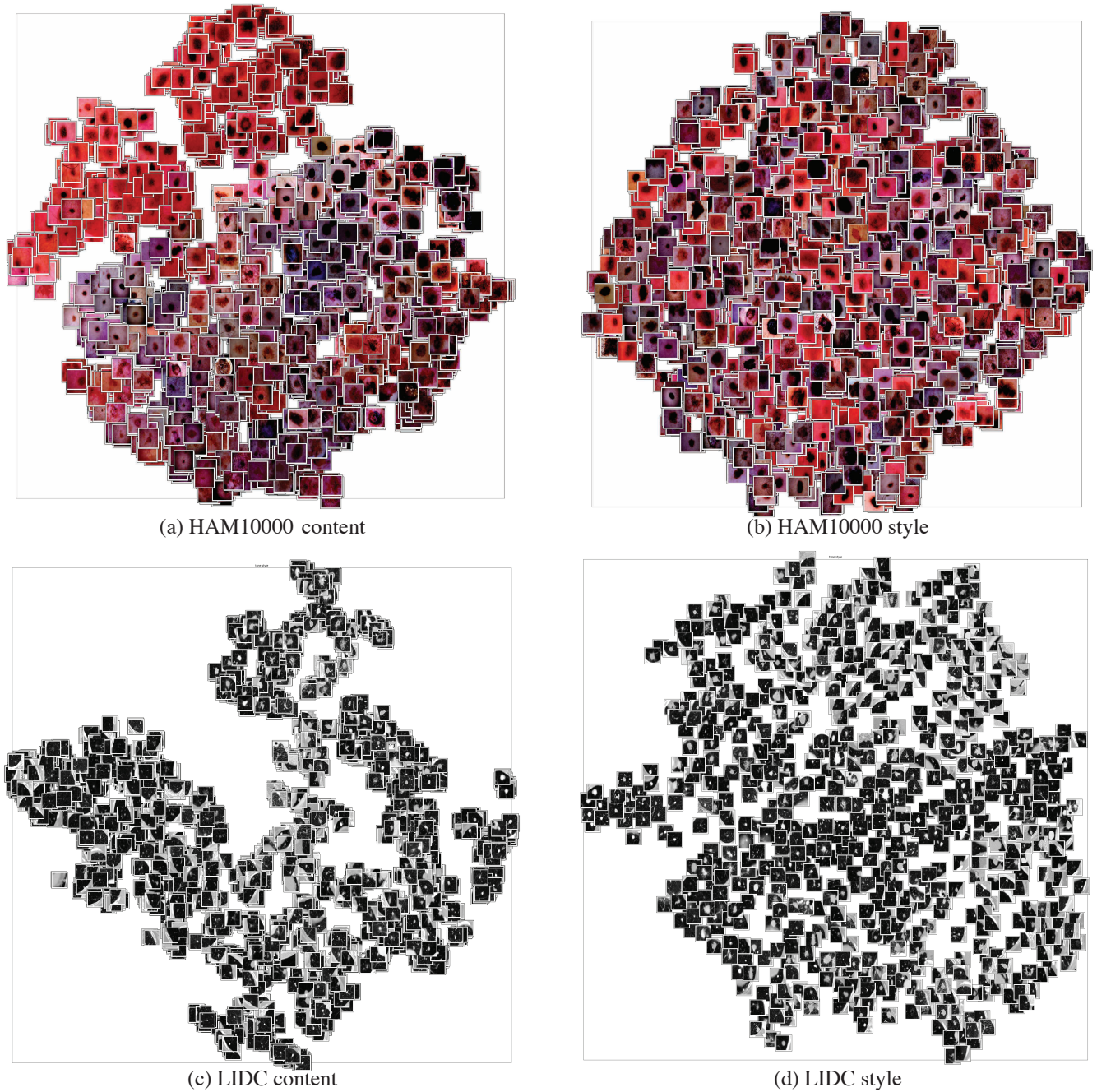}
	\caption{t-SNE plots of the inferred content (a and c) and style (b and d) codes for LIDC and HAM10000. In the case of HAM10000, the content variable focuses mostly on color and size whereas the style variable focuses on lesion location and orientation. In the case of LIDC, the content variable focuses mostly on nodule size whereas the style variable focuses on background information.
	}
	\label{fig:t-sne}
	\vspace{-3mm}%Put here to reduce too much white space after your table 
\end{figure*}

\paragraph{{\bf \textit{t-SNE plots for style and content latent spaces}}}
To better understand how DRAI clusters style and content information, we provide t-SNE \citep{van2008visualizing} plots of the style and content variables for both datasets in 2 dimensions. 
As seen in Figure~\ref{fig:t-sne}, for LIDC, the style clusters similar backgrounds together while the content focuses more on the size of the nodules. Whereas for HAM10000, the style focuses more on the shape, orientation and position of the lesion while the content focuses more on size of the lesions and color.

\input{ablation_lidc}
\input{ablation_ham}

%\subsubsection{Importance of Feature cycle-consistency}
%\subsubsection{Importance of content-style information minimization}
%\subsubsection{Importance of self-supervised regularization}
%Figures~\ref{fig:ablation-ham},
%\subsection{Does supervised content help disentanglement?}
\subsection{Implementation details}
In this section, we provide the important implementation details of DRAI. Firstly, to reduce the risk of information leak between style and content, we use completely separate encoders to infer the two variables. For the same reason, the dual adversarial discriminators are also implemented separately for style and content. The data augmentation includes random flipping and cropping. To enable self-supervised regularization, each batch is trained twice, first with the original images and then with the transformed batch. The transformations include rotations of 90, 180, and 270 degrees, as well as horizontal and vertical flipping. %The transformation process is controlled so that each input transforms exactly once with any one of the transformations described. 
%The transformed batch is introduced to enable the self-supervised regularization. 
%In addition, GRL is used to minimize the similarity between the inferred style features of the original and transformed batch.
%We implement GRL with constant reverse gradient coefficient, in other words, the gradient in the backward pass is always multiplied by $-1$. Furthermore, we add KL divergence for Gaussian latent variable in the loss function. Lastly, 
LSGAN (Least Square GAN)~\citep{mao2017least} loss is used for all GAN generators and discriminators, while $\ell_1$ loss is used for the components related to disentanglement constraints, \ie, GRL strategy and self-supervised regularization. In general, we found that ``Image cycle-consistency'' and ``Latent code cycle-consistency" objectives improve the stability of training. This is evident by DRAI achieving lower prediction intervals (\ie, standard deviation across multiple runs with different seeds) in our experiments. 

We did not introduce any coefficients for the loss components in Equation~(\ref{eq:full-obj}) since other than the KL terms, they were all relatively on the same scale. As for the KL co-efficients $\lambda$, we tried multiple values and qualitatively evaluated the results. Since the model was not overly sensitive to KL, we used a coefficient of $1$ for all KL components.

All models including the baselines are implemented in TensorFlow \citep{tensorflow2015-whitepaper} version 2.1, and the models are optimized via Adam \citep{kingma2015adam} with initial learning rate $1e^{-5}$. For IS and FID computation, we fine-tune the inception model on a 5 way  classification on nodule size for LIDC and a 7 way classification on lesion type for HAM10000. FID and IS are computed over a set of 5000 generated images.
%[ Into the appendix?
%- c/s dims for lidc/ham
%- loss function coefficients [To what extends?]
%- kl coefficients ... etc.]

%% file: lidc.tex
 \begin{table}[t!]
\caption{ Comparison of image generation metrics on LIDC dataset of single and double variable baselines}
\centering
\resizebox{\columnwidth}{!}{
\begin{tabular}{l*{3}{c}}
\toprule
Method \multirow{2}*{ }&\multicolumn{3}{c}{LIDC}\\
\cmidrule{2-4}
&FID($\downarrow$) &IS($\uparrow$) &CGAcc($\uparrow$) \\ \midrule
cInfoGAN	& $0.283 \pm 0.06$	& $1.366 \pm 0.02$ & $0.740 \pm 0.02$\\
cAVAE	& $0.181 \pm 0.03$	& $1.424 \pm 0.01$ & $0.809 \pm 0.02$\\
D-cInfoGAN	& $0.333 \pm 0.06$	& $1.342 \pm 0.09$ & $0.645 \pm 0.04$\\
D-cAVAE	& $0.378 \pm 0.03$	& $1.371 \pm 0.04$ & $0.597 \pm 0.07$\\
DAI~\citep{lao2019dual}	        & $0.106 \pm 0.02$	& $1.423 \pm 0.05$ & $0.773 \pm 0.03$\\
DRAI	    & $0.089 \pm 0.02$	& $1.422 \pm 0.03$ & $0.773 \pm 0.02$\\
\bottomrule
\end{tabular}
 }
\label{tab:FID-IS-LIDC}
%\vspace*{-0.5cm}
\end{table}

%% file: _ham.tex
\begin{table}[t!]
\caption{Comparison of image generation metrics on HAM10000 dataset of single and double variable baselines.}
\centering
\resizebox{\columnwidth}{!}{
\begin{tabular}{l*{3}{c}}
\toprule
Method \multirow{2}*{ }&\multicolumn{3}{c}{HAM}\\
\cmidrule{2-4}
&FID($\downarrow$) &IS($\uparrow$) &CGAcc($\uparrow$) \\ \midrule
cInfoGAN	& $1.351 \pm 0.33$	& $1.326 \pm 0.03$ & $0.647 \pm 0.04$\\
cAVAE	& $3.566 \pm 0.56$	& $1.371 \pm 0.01$ & $0.651 \pm 0.02$\\
D-cInfoGAN	& $1.684 \pm 0.42$	& $1.449 \pm 0.03$ & $0.654 \pm 0.06$\\
D-cAVAE	& $4.893 \pm 0.99$   & $1.321 \pm 0.01$ & $0.578 \pm 0.01$\\				
DAI~\citep{lao2019dual}       	& $1.327 \pm 0.06$	& $1.304 \pm 0.01$ & $0.656 \pm 0.02$\\
DRAI	    & $1.224 \pm 0.05$	& $1.300 \pm 0.01$ & $0.630 \pm 0.02$\\
\bottomrule
\end{tabular}
}
\label{tab:FID-IS-HAM}
%\vspace*{-0.5cm}
\end{table}

%% file: disentangle_err.tex
\begin{table}[t!]
\caption{Content-style disentanglement evaluation using CIFC error.}
\centering
\begin{adjustbox}{width=0.8\columnwidth}
\begin{tabular}{l*{2}{c}}
\toprule
Method \multirow{2}*{ }&\multicolumn{2}{c}{CIFC($\downarrow$)}\\
\cmidrule{2-3}& HAM10000 & LIDC\\
\midrule
D-cInfoGAN  & $1.201\pm0.17$  &  $1.625\pm0.11$\\
D-cAVAE & $1.354\pm0.03$  &  $1.944\pm0.02$\\
DAI~\citep{lao2019dual}           & $0.256\pm0.01$  &  $1.096\pm0.28$\\
DRAI       & $0.210\pm0.01$  &  $0.456\pm0.06$\\

\bottomrule
\end{tabular}
\end{adjustbox}
\label{tab:disentangle}
%\vspace*{-0.5cm}
\end{table}

%% file: retreival.tex
\begin{table}[t!]
\caption{ Quantitative evaluation on the top 3 nearest neighbors for content based image retrieval on LIDC dataset.}
\centering
\resizebox{\columnwidth}{!}{
\begin{tabular}{l*{2}{c}}
\toprule
Method & disagreement divergence($\downarrow$) & label-overlap($\uparrow$) \\
Pixel~Matched   &$2.42$    & $35.27$    \\
\midrule
D-cInfoGAN       & $2.566\pm0.04$ & $33.850\pm1.19$ \\
D-cAVAE      & $2.961\pm0.08$ & $35.010\pm0.96$ \\
\midrule
DAI+selfReg+MIReg        & $1.826\pm0.05$	& $52.325\pm3.87$ \\
DAI+featureCycle           & $1.458\pm0.11$	& $59.173\pm3.16$ \\
DAI+MIReg           & $1.928\pm0.26$	& $54.909\pm1.82$ \\
DAI+selfReg        & $1.791\pm0.26$	& $56.589\pm3.59$ \\
\midrule
DAI~\citep{lao2019dual}                 & $1.687\pm0.32$	& $54.780\pm5.02$ \\
DRAI     & $1.575\pm0.04$	& $53.359\pm2.94$ \\

\bottomrule
\end{tabular}
}

\label{tab:retrieval}
%\vspace*{-0.5cm}
\end{table}

%% file: ablation_lidc.tex
%\begin{table*}[h!]
\begin{table*}[ht]
\caption{Quantitative ablation study on LIDC dataset}
\centering
\begin{adjustbox}{width=\textwidth}
\setlength{\tabcolsep}{2em}
\begin{tabular}{l*{4}{c}}
\toprule
Method \multirow{2}*{ }&\multicolumn{4}{c}{LIDC}\\
\cmidrule{2-5}
&FID($\downarrow$) &IS($\uparrow$) &CGAcc($\uparrow$) & CIFC($\downarrow$) \\ \midrule
DAI~\cite{lao2019dual}	            &  $0.106\pm0.02$	&  $1.423\pm0.052$ &  $0.773\pm0.031$  &  $1.096\pm0.284$\\
DRAI=DAI+selfReg+MIReg+featureCycle	&  $0.089\pm0.02$	&  $1.422\pm0.030$ &  $0.773\pm0.016$  &  $0.456\pm0.069$\\
DAI+selfReg+MIReg	    &  $0.176\pm0.06$	&  $1.433\pm0.018$ &  $0.760\pm0.015$  &  $0.554\pm0.185$\\
DAI+featureCycle	        &  $0.221\pm0.07$	&  $1.383\pm0.039$ &  $0.708\pm0.037$  &  $0.913\pm0.074$\\
DAI+MIReg	        &  $0.154\pm0.04$	&  $1.411\pm0.028$ &  $0.746\pm0.041$  &  $0.747\pm0.226$\\
DAI+selfReg	    &  $0.208\pm0.05$	&  $1.433\pm0.033$ &  $0.780\pm0.014$  &  $0.781\pm0.203$\\
\bottomrule
\end{tabular}

\end{adjustbox}
\label{tab:ablation_lidc}
%\vspace*{-0.5cm}
\end{table*}

%% file: ablation_ham.tex
\begin{table*}[ht]
\centering
\caption{Quantitative ablation study on HAM10000 dataset.}
\begin{adjustbox}{width=\textwidth}
\setlength{\tabcolsep}{2em}

\begin{tabular}{l*{4}{c}}
\toprule
Method \multirow{2}*{ }&\multicolumn{4}{c}{HAM}\\
\cmidrule{2-5}
&FID($\downarrow$) &IS($\uparrow$) &CGAcc($\uparrow$) & CIFC($\downarrow$) \\ \midrule
DAI~\citep{lao2019dual}                &$1.327\pm 0.06$ & $0.656\pm0.02$ & $1.304\pm0.01$ & $0.256\pm0.01$\\
DRAI=DAI+selfReg+MIReg+featureCycle    &$1.224\pm0.050$ & $0.618\pm0.011$ & $1.300\pm0.01$ & $0.210\pm0.01$\\
DAI+selfReg+MIReg        &$1.350\pm0.12$ & $0.683\pm0.02$ & $1.299\pm0.01$ & $0.233\pm0.01$\\
DAI+featureCycle           &$1.367\pm0.12$ & $0.690\pm0.01$ & $1.296\pm0.01$ & $0.311\pm0.01$\\
DAI+MIReg          &$1.298\pm0.12$ & $0.647\pm0.04$ & $1.290\pm0.02$ & $0.228\pm0.01$\\
DAI+selfReg       &$1.347\pm0.14$ & $0.653\pm0.03$ & $1.295\pm0.01$ & $0.219\pm0.04$\\
\bottomrule
\end{tabular}
\end{adjustbox}
\label{tab:ablation_ham}
%\vspace*{-0.5cm}
\end{table*}

%% file: related_work.tex
%Related work
\paragraph{Connection to other conditional GANs in medical imaging}
While adversarial training has been used extensively in the medical imaging domain, most work uses adversarial training to improve image segmentation and domain adaptation. 
The methods that use adversarial learning for image generation can be divided into two broad categories;
the first group are those which use image-to-image translation as a proxy to image generation. These models use an image mask as the conditioning factor, and the generator generates an image which respects the constraints imposed by the mask~\cite{jin2018ct,guibas2017synthetic,costa2017towards,costa2017towards,mok2018learning}. \citet{jin2018ct} condition the generative adversarial network on a 3D mask, for lung nodule generation. In order to embed the nodules within their background context, the GAN is conditioned on a volume of interest whose central part containing the nodule has been erased.
A favored approach for generating synthetic fundus retinal images is to use vessel segmentation maps as the conditioning factor. 
\citet{guibas2017synthetic} uses two GANs in sequence to generate fundus images. The first GAN generates vessel  masks, and in stage two, a second GAN is trained to generate fundus retinal images from the vessel masks of stage one. 
%a mask and the other uses the generated mask to produce an image.
\citet{costa2017towards} first use a U-Net based model to generate vessel segmentation masks from fundus images. An adversarial image-to-image translation model is then used to translate the mask back to the original image. 

In \citet{mok2018learning} the generator is conditioned on a brain tumor mask and generates brain MRI. To ensure correspondence between the tumour in the generated image and the mask, they further forced the generator to output the tumour boundaries in the generation process. \citet{bissoto2018skin} uses the semantic segmentation of skin lesions and generate high resolution images. Their model combines the pix2pix framework~\cite{isola2017image} with multi-scale discriminators to iteratively generate coarse to fine images.

While methods in this category give a lot of control over the generated images, the generator is limited to learning domain information such as low level texture and not higher level information such as shape and composition. Such information is presented in the mask which requires an additional model or an expert has to manually outline the mask which can get tedious for a lot of images. 

The second category of methods are those which use high level class information in the form of a vector as the conditioning factor. 
~\citet{hu2018prostategan} takes Gleason score vector as input to the conditional GAN to generate synthetic prostate diffusion imaging data corresponding to a particular cancer grade. \citet{baur2018generating} used a progressively growing model to generate high resolution images of skin lesions. 

As mentioned in the introduction one potential pitfall of such methods is that by just using the class label as conditioning factor, it is hard to have control over the nuances of every class. While our proposed model falls within this category, our inference mechanism allows us to overcome this challenge by using the image data itself to discover factors of variation corresponding to various nuances of the content.

\paragraph{Disentangled representation learning}
In the literature, disentanglement of style and content is primarily used for domain translation or domain adaptation. Content is defined as domain agnostic information shared between the domains, while style is defined as domain specific information. The goal of disentanglement to preserve as much content as possible and to prevent leakage of style from one domain to another. 
\citet{gonzalez2018image} used adversarial disentanglement for image to image translation. In order to prevent exposure of style from domain A to domain B, a Gradient Reversal Layer (GRL) is used to penalize shared information between the generator of domain B and style of domain A. In contrast, our proposed DRAI, uses GRL to minimize the shared information between style and content. 
In the medical domain, \citet{yang2019domain} aim to disentangle anatomical information and modality information in order to improve on a downstream liver segmentation task. %The disentanglement is done by ...

\citet{ben2019improving} used adversarial learning to infer content agnostic features as style. Intuitively their method is similar to using GRL to minimize leakage of content information into a style variable. However, while ~\cite{ben2019improving} prevents leakage of content into style, it does not prevent the reverse effect which is leakage of style into content and thus does not guarantee disentanglement. 

\citet{yang2019unsupervised} use disentangle learning of modality agnostic and modality specific features in order to facilitate cross-modality liver segmentation. They use a mixture of adversarial training and cycle consistency loss to achieve disentanglement. The cycle-consistency  component is used for in-domain reconstruction and the adversarial component is used for cross-domain translation. The two components encourage the disentanglement of the latent space, decomposing it into modality agnostic and modality specific sub-spaces.  

To achieve disentanglement between modality information and anatomical structures in cardiac MR images, \citet{chartsias2019disentangled} use an autoencoder with two encoders: one for the modality information (style) and another for anatomical structures (content). They further impose constraints on the anatomical encoder such that every encoded pixel of the input image has a categorical distribution. As a result, the output of the anatomical encoder is a set of binary maps corresponding to cardiac substructures. 

Disentangled representation learning has also been used for denoising of medical images. In \citet{liao2019adn}, Given artifact affected CT images, metal-artifact reduction (MAR) is performed by disentangling the metal-artifact representations from the underlying CT images.  

\citet{sarhan2019} use $\beta$-$\text{TCAV}$ \cite{NIPS2018_7527} to learn disentangled representations on an adversarial variation of the VAE. Their proposed model differs fundamentally from our work; its is a single variable model, without a conditional generative process, and does not infer separate style and content information.

\citet{Garcia12019Towards} used ALI (single variable) on structured MRI to discover regions of the brain that are involved in Autism Spectrum Disorder (ASD).

In contrast to previous work, %that use disentangled learning for denoising or domain translation,
we use style-content disentanglement to control features for conditional image generation. To the best of our knowledge this is the first time such attempt has been made in the context of medical imaging. 
%\paragraph{weakly supervised representation learning}
%\paragraph{Information bottleneck}
\paragraph{Gradient Reverse Layer (GRL)}
Introduced in \citet{GaninUAGLLML15}, GRL uses the identity function in the forward pass, but reverses the sign of the gradient in the backward pass. GRL was initially intended for domain adaptation with the goal of learning domain invariant representations. Since then, GRL has been used extensively in various domain adaptation approaches~\cite{zhang2019bridging, jiang2020implicit, lao2020continuous}. For image translation, \citet{gonzalez2018image} used a model based on VAE-GAN and GRL to disentangle the attributes of paired data into shared and exclusive representations. \citet{raff2018gradient} used GRL for {\it fairness} with the goal of learning a model whose outcome is invariant to sensitive attributes such as ethnicity or gender. It is worth mentioning that some fairness and invariant representation learning methods, rely on adversarial training which in spirit is similar to using GRL \cite{kim2019learning,madras2018learning}. 

Different from previous work, we introduce a novel use case for GRL to explicitly disentangle the inferred style and content information.

\paragraph{Connection to CycleGAN approaches}
\citet{zhu2017unpaired} introduced CycleGAN as an image-translation (domain-translation) model which preserves the content and changes the style from one domain to another domain. At test time, the mapping between the domains is deterministic and one-to-one.~\citet{almahairi2018augmented} augmented CycleGAN with latent variables which allowed many to many mapping between the two domains. 
In this paper we do not tackle the problem of image domain translation and rather focus on conditional generation of desired content and style through dual adversarial inference. In our approach, the cycle-consistency is used to stabilize the adversarial inference. That being said, it would be possible to apply our framework to image domain translation problem as well. We leave exploration in this direction to future work.

%% file: conclusion.tex
We introduce DRAI, a frame work for generating synthetic medical images which allows control over the style and content of the generated images. 

DRAI uses adversarial inference together with conditional generation and disentanglement constraints to learn content and style variables from the dataset. 
We compare DRAI quantitatively and qualitatively with multiple baselines and show its superiority in image generation in terms of quality, diversity and style-content disentanglement. 
Through ablation studies and comparisons with DAI~\cite{lao2019dual}, we show the impact of  imposing the proposed disentanglement constraints over the content and style variables. 

%Compared with the baselines, DRAI achieves better disentanglement score on two publically available medical imaging datasts.    
It is important to note that DRAI learns style in a completely unsupervised fashion, allowing for its practical use case in real world scenarios where information contributing to style is not available. Also, contrary to previous methods that assume a high-level conditioning vector for generation, in addition to using supervised learning to learn the content, our framework uses unsupervised learning to discover factors of variation not present in the conditioning vector. 

The proposed model has a wide range of potential applications from improving the training of deep models by means of data augmentation to generating rare image cases for training of medical personnel.  

In future work we would like to explore  DRAI's ability to improve generalization of supervised learning through conditional synthetic image generation. It is also interesting to explore whether DRAI's ability to disentangle style and content can help to learn better representations of under-represented subgroups.

%% file: acknowledgement.tex
\section{Acknowledgement}
We would like to thank Joseph Viviano, Xinag Jiang, Lisa Di Jorio, Francis Dutil, Tess Berthier and Louis-\'Emile Robitaille for constructive discussions and feedback, and Tanya Nair for proof reading the paper.